\documentclass{ieeeaccess}
\usepackage{cite}
\usepackage{amsmath,amssymb,amsfonts}
\usepackage{algorithm}
\usepackage{algorithmicx}
\usepackage{algpseudocode}
\usepackage{csquotes}
\usepackage[caption=false]{subfig}
\algnewcommand{\LineComment}[1]{\State \(\triangleright\) #1}
\usepackage{graphicx}
\usepackage{soul}
\usepackage{textcomp}
\usepackage{multirow}

\newtheorem{definition}{Definition}[section]
\newtheorem{theorem}{Theorem}[section]

\def\BibTeX{{\rm B\kern-.05em{\sc i\kern-.025em b}\kern-.08em
    T\kern-.1667em\lower.7ex\hbox{E}\kern-.125emX}}
\begin{document}
\history{Date of publication xxxx 00, 0000, date of current version xxxx 00, 0000.}
\doi{10.1109/ACCESS.2017.DOI}

\title{Successive Point-of-Interest Recommendation with Local Differential Privacy}
\author{\uppercase{Jong Seon Kim}\authorrefmark{1},
\uppercase{Jong Wook Kim\authorrefmark{2}, \IEEEmembership{Member, IEEE} and Yon Dohn Chung}\authorrefmark{1},
\IEEEmembership{Member, IEEE}}
\address[1]{Department of Computer Science and Engineering, Korea University, Seoul 02841, South Korea}
\address[2]{Department of Computer Science, Sangmyung University, Seoul 03016, South Korea}
\tfootnote{This research was partially supported by the MSIT(Ministry of Science and ICT), Korea, under the ICT Creative Consilience program(IITP-2021-2020-0-01819) supervised by the IITP(Institute for Information \& communications Technology Planning \& Evaluation), and the National Research Foundation of Korea(NRF) grant funded by the Korea government(MSIT) (No. NRF-2020R1A2C2013286).}

\markboth
{J.S.Kim \headeretal: Successive Point-of-Interest Recommendation with Local Differential Privacy}
{J.S.Kim \headeretal: Successive Point-of-Interest Recommendation with Local Differential Privacy}

\corresp{Corresponding author: Yon Dohn Chung (e-mail: ydchung@korea.ac.kr).}

\begin{abstract}
A point-of-interest (POI) recommendation system performs an important role in location-based services because it can help people to explore new locations and promote advertisers to launch advertisements at appropriate locations. The existing POI recommendation systems require raw check-in history of users, which might cause location privacy violations. Although there have been several matrix factorization (MF) based privacy-preserving recommendation systems, they can only focus on user-POI relationships without considering the human movements in check-in history. To tackle this problem, we design a successive POI recommendation framework with local differential privacy, named SPIREL. SPIREL uses two types of information derived from the check-in history as input for the factorization: a transition pattern between two POIs and the visit counts of POIs. We propose a novel objective function for learning the user-POI and POI-POI relationships simultaneously. We further integrate local differential privacy mechanisms in our proposed framework to prevent potential location privacy breaches. Experiments using four public datasets demonstrate that SPIREL achieves better POI recommendation quality while accomplishing stronger privacy protection.
\end{abstract}


\titlepgskip=-15pt

\maketitle

\section{Introduction}
\label{introduction}

Smartphones have become an integral part of our everyday lives. In particular, smartphones have resulted in people sharing their daily check-in experiences through social network services, such as Facebook, Foursquare, and Instagram. Through these check-in data, it is possible to study the online activities, physical movements, and preferences on the points-of-interest (POI) of users. Accordingly, various location-based services (LBSs) utilize the check-in data to provide the best experiences for their services. Among the various tasks in LBSs, POI recommendation has attracted considerable attention in recent years \cite{cheng2013you,liu2013personalized,lian2014geomf,chang2018content}.

Predicting the subsequent location of a mobile user is important as it can help them to explore interesting and unvisited places. To recommend new POIs for users, most of the recommendation methods leverage the check-in history of users. As the check-in history embeds rich sequential patterns, it is intuitive to utilize it to decide on the next location of users. However, collecting the check-in data poses serious privacy concerns. For example, a previous  study \cite{de2013unique} demonstrated that only four successive location points are sufficient to uniquely identify each individual. Alongside this kind of identity disclosure problem, it is also possible to infer private information through the visited places. Another study \cite{jung2020too} showed that they could infer sensitive information (e.g., sexual preferences or religious affiliations) through the check-in data collected for tracing people infected with COVID-19.

The notion of local differential privacy (LDP) has attracted considerable attention in recent years from many industries owing to its rigorous and provable privacy guarantees. In the LDP setting, each user randomizes the original data in his/her device and sends the perturbed data to the server. As the original data remain within the devices, the server cannot infer the sensitive information about users, regardless of the background knowledge. Accordingly, many global IT companies, including Google\cite{erlingsson2014rappor}, Apple\cite{apple2014learning}, Microsoft\cite{ding2017collecting}, and Samsung\cite{nguyen2016collecting}, have already adopted the LDP mechanisms for collecting data from their clients.

Several earlier works adopted differential privacy (DP) in the recommendation system for preserving the privacy of users \cite{hua2015differentially, zhang2018probabilistic, shin2018privacy}. Hua et al.\cite{hua2015differentially} proposed a recommendation system based on the centralized differential privacy (CDP) setting. However, the server is assumed to be trustworthy in the CDP setting, which means the data is collected without any privacy guarantee. Zhang et al.\cite{zhang2018probabilistic} proposed a recommendation system based on a personalized CDP model. Finally, Shin et al.\cite{shin2018privacy} proposed a recommendation system under LDP. In their work, the LDP mechanism was utilized by involving a stochastic gradient descent (SGD) step for updating the latent factors, where each user sends perturbed gradients to the recommendation server.

The above privacy-preserving recommendation systems have two limitations in their solutions. First, \cite{hua2015differentially} and \cite{zhang2018probabilistic} assumed that the recommenders should have a trust relationship with the users. That is, the check-in data is submitted in its original form. Second, all the studies \cite{hua2015differentially, zhang2018probabilistic, shin2018privacy} are developed on single-domain (SD) recommendation systems that focus on the user-item relationship only. If we build a POI recommendation system based on these methods, the system will highly recommend unattractive POI candidates without considering the sequential characteristics of POIs.

In this paper, we propose a novel privacy-preserving POI recommendation framework called SPIREL (\textbf{S}uccessive \textbf{P}O\textbf{I} \textbf{RE}commendation with \textbf{L}DP). SPIREL recommends the next POI candidates while considering the sequential preferences on POIs. To do this, SPIREL uses matrix factorization (MF) \cite{koren2009matrix} among many collaborative filtering (CF) techniques. SPIREL exploits two types of implicit interactions from the check-in history of the user. First, to identify the sequential preferences on POIs, SPIREL uses transition patterns modeled with a first-order Markov chain. Specifically, each user is required to record a movement between two successive POIs. Second, SPIREL further extracts the visit counts for each POI from the check-in history of the user. In summary, SPIREL jointly learns the user-POI and POI-POI relationships based on the visit counts and transition patterns, respectively.

The state-of-the-art privacy-preserving MF-based recommendation system under LDP \cite{shin2018privacy} is designed for explicit feedback data, where users denote their "like" or "dislike" through a fixed scale rating. Accordingly, \cite{shin2018privacy} focuses on reducing the amount of noise added on gradients without modeling the user preferences.
On the other hand, lacking explicit ratings is the main problem in POI recommendations. One way is to guess the preferences through the visit counts that only reveal positive interactions between users and POIs. However, visit counts alone cannot deeply explain the users' preferences for the next POIs. Thus, we measure sequential preferences on POIs across users by mapping the estimated frequency of transition patterns to a confidence score. Moreover, we design a novel objective function to integrate the confidence scores, so the preference for the next POI is modeled as an inner product between visit counts and the confidence scores.

The main contributions of our work can be summarized as follows.

\begin{enumerate}
    \item To map the implicit feedback to the sequential preferences, we sample a single POI-POI transition pattern from each user and estimate the frequency of sampled transition patterns under LDP. Then, we transfer the estimated frequency of transition patterns into a confidence score to model the POI transition preferences.
    \item During the learning process, our proposed framework simultaneously factorizes both user-POI and POI-POI matrices. To achieve this, we developed a novel objective function and an optimization framework under LDP. Note that the existing methods are SD recommendation systems that can factorize only the user-POI matrix, where they have limits in integrating additional knowledge from other domains.
    \item In our framework, the users do not have to disclose their current location to receive the next POI candidates. Considering that most LBSs need the query locations of users, SPIREL is an end-to-end privacy-aware framework for successive POI recommendations.
    \item We conducted extensive experiments using four public datasets and demonstrated that SPIREL improves the accuracy and quality of recommendations compared to the previous privacy-preserving recommendation systems.
\end{enumerate}

The remainder of this paper is organized as follows. In Section \ref{preliminaries}, we explain the background knowledge. In Section \ref{problem definition}, we formally define the problem and describe the challenges. In Section \ref{proposed method}, we propose our SPIREL framework for successive POI recommendation. Section \ref{evaluation} demonstrates the performance of SPIREL on public datasets. Section \ref{related work} reviews the related work. Finally, in Section \ref{conclusion}, we conclude this paper.

\section{Preliminaries}
\label{preliminaries}

\begin{table}[ht]
\renewcommand{\arraystretch}{1.3}
\caption{Notations}
\label{table1}
\centering
\begin{tabular}{|c||c|}
\hline
Notation & Meaning\\
\hline
\hline
$m$ & number of users\\
\hline
$n$ & number of POIs\\
\hline
$d$ & size of latent factors\\
\hline
$\mathbf{u_{i}} \in \mathbb{R}^{d}$ & profile vector of user $i$\\
\hline
$\mathbf{v_{j}} \in \mathbb{R}^{d}$ & profile vector of POI $j$\\
\hline
$U \in \mathbb{R}^{m \times d}$ & user profile matrix \\
\hline
$V \in \mathbb{R}^{n \times d}$ & POI profile matrix\\
\hline
$P \in \mathbb{R}^{m \times n}$ & user-POI matrix\\
\hline
$Q \in \mathbb{R}^{n \times n}$ & POI-POI matrix\\
\hline
$P_{i,*}$ & $i$th row vector of matrix $P$\\
\hline
$P_{*,j}$ & $j$th column vector of matrix $P$\\
\hline
\end{tabular}
\end{table}

\subsection{Matrix Factorization}
\label{matrix factorization}

MF is used as a CF algorithm in recommendation systems \cite{koren2009matrix}. In recent years, owing to its accurate prediction, many industries have adopted it for personalized advertisement targeting. MF decomposes the user-item matrix into two smaller matrices to discover the unobserved relationship between users and items. Each decomposed matrix embeds the latent factors of the user/item, which simplifies the complicated user/item characteristics. Then, by multiplying the two latent matrices, we can predict the unobserved user-item relationships. In this study, we assume that the latent factors of the item represent the characteristics of POIs.

The major challenge involved in MF is the optimization process for finding the mapping between users and items. Let us assume that the elements of the user-item matrix indicate the observed preference scores of users on items. Then, the objective of MF is to determine the latent factors of the user and item, whose product becomes similar to the observed preference scores. To be specific, MF attempts to reduce the error between the observed preference scores and the predicted scores calculated over an inner product of the two latent factors. While minimizing these errors, the latent factors are fitted to uncover the preference scores of items that are not rated.

Formally, MF is defined as follows. Suppose there are $m$ users and $n$ items. We denote $r_{ij}$ as the preference score of user $i$ for item $j$. Then, $r_{ij}$ can be approximated by calculating the inner product $r_{ij} \approx \mathbf{u_{i}}^{\intercal}\mathbf{v_{j}}$. The objective of MF is to minimize the error between $r_{ij}$ and $\mathbf{u_{i}}^{\intercal}\mathbf{v_{j}}$. In Table \ref{table1}, we have listed the set of notations used throughout this paper. Unless otherwise stated, we assume that all vectors are column vectors.

Generally, as users rate only a small set of items, we have a considerably limited number of observed feedback. Thus, the user-item matrix $P$ is significantly sparse, which implies that most of the elements in $P$ are unknown. Therefore, many of the recent studies performed optimization with only the observed scores while avoiding overfitting by introducing a regularization term. Specifically, MF attempts to minimize the following objective function.

\begin{equation} 
\label{eq1}
\mathcal{L} = \sum_{{(i,j)} \in P} (r_{ij}-\mathbf{u_{i}}^{\intercal}\mathbf{v_{j}})^{2} + \lambda(\sum_{i=1}^{m}||\mathbf{u_{i}}||^{2} + \sum_{j=1}^{n}||\mathbf{v_{j}}||^{2})
\end{equation}

Here, the user profile vector $\mathbf{u_{i}}$ and item profile vector $\mathbf{v_{j}}$ are represented by a $d$-dimensional vector. Furthermore, $\lambda$ is a regularizer used for avoiding the overfitting problem. In summary, by minimizing the mean square error over the known preference scores, we can predict the unobserved user-item relationships.

There are primarily two optimization techniques to minimize the objective function: (1) SGD and (2) alternating least square (ALS). SGD first predicts $r_{ij}$ and computes associated prediction error for each observed case. Then, SGD derives the gradient from the corresponding latent factors $\mathbf{u_{i}}$ and $\mathbf{v_{i}}$, which indicates the direction of the greatest rate of increase of the objective function. The gradient is multiplied with a learning rate $\gamma$, which determines the change in the profile vector with respect to the gradient. Subsequently, SGD performs steps in the inverse direction of the gradient to minimize the objective function. We describe the update rule of the user profile vector below (the item profile vector can be updated similarly).

\begin{equation} 
\label{eq2}
\nabla_{\mathbf{u_{i}}}\mathcal{L} = \sum_{{(i,j)} \in P} -2\mathbf{v_{j}}(r_{ij}-\mathbf{u_{i}}^{\intercal}\mathbf{v_{j}}) + 2\lambda \mathbf{u_{i}}
\end{equation}

\begin{equation} 
\label{eq3}
\mathbf{u_{i}} = \mathbf{u_{i}} - \gamma \cdot  \nabla_{\mathbf{u_{i}}}\mathcal{L}
\end{equation}

Next, we briefly introduce ALS. It is difficult to optimize $\mathbf{u_{i}}$ and $\mathbf{v_{j}}$ jointly, as minimizing Equation \ref{eq1} is a non-convex problem. One way to solve this problem is to minimize the equation by fixing $\mathbf{u_{i}}$ and $\mathbf{v_{i}}$ in an alternative manner. For example, consider the item profile vector as a constant and calculate the derivative of Equation \ref{eq1} with respect to $\mathbf{u_{i}}$. Then, set the derivative to zero and solve the quadratic equation to obtain the update rule of $\mathbf{u_{i}}$ as follows.

\begin{equation} 
\label{eq4}
\begin{split}
\frac{\partial \mathcal{L}}{\partial \mathbf{u_{i}}} &= -2\sum_{j}  (r_{ij}-\mathbf{u_{i}}^{\intercal}\mathbf{v_{j}})\mathbf{v_{j}}^{\intercal} + 2\lambda \mathbf{u_{i}}^{\intercal} \\
0 &= -(P_{i,*} - \mathbf{u_{i}}^{\intercal}V^{\intercal})V + \lambda \mathbf{u_{i}}^{\intercal} \\
\mathbf{u_{i}}^{\intercal} &= P_{i,*}V(V^{\intercal}V + \lambda I)^{-1}
\end{split}
\end{equation}

In terms of convergence, ALS is faster than SGD and generally completed the learning process in the first twenty iterations \cite{tan2016faster
}. SGD requires more iterations to converge; however, SGD can reach a lower mean square error than that of ALS. We combine both SGD and ALS to minimize the loss function of SPIREL, and the combined method can integrate the advantages of both techniques and complement the shortcomings of them.

\subsection{Transfer Learning}
\label{transfer learning}

CF algorithms rely on different types of input data to estimate the preference scores of users on items. A common approach is to extract the preference scores directly from explicit feedback (e.g., movie rating) or to estimate the scores from implicit feedback (e.g., purchase history). Generally, using explicit feedback leads to more accurate results as a user reveals the preferences for items directly. In our framework, the input data is a check-in history that provides only positive samples (i.e., the places the users had visited), which implicitly reflects the preferences of POIs.

Most of the recommendation systems are \textit{SD recommendation systems}, which only focus on one domain while ignoring various interactions between users and items. In several real-world applications, user-item relationships are recorded over time as event sequences; moreover, the check-in history is one such example. Accordingly, when directly building the POI recommendation system over the SD recommendation system, the natural idea is to utilize the visit counts to learn interests of users on POIs. However, such methods easily suffer from the data sparsity issue as the users typically visit only a few locations.

In recent years, the notion of \textit{cross-domain (CD) recommendation} \cite{khan2017cross}, which leverages additional knowledge learned from other domains to improve the target domain recommendation task, has emerged. The CD recommendation systems are known to overcome the data sparsity issue in the SD recommendation system. One of the widely studied approaches is \textit{transfer learning}. Previous studies \cite{pan2010transfer, pan2011transfer, pan2013transfer} introduced the transfer learning approach to MF. The primary idea of using transfer learning within MF is to share the latent factors between the source domain and the target domain, assuming that the shared latent factors have similar characteristics.

In our work, to assist recommending the next POIs from the user-POI domain modeled with the visit counts, we integrated the confidence score for each POI transition pattern derived from the LDP mechanism. Consequently, we transferred the knowledge learned from POI transition patterns to a user-POI domain, assuming that the latent factors of the POI can bridge the static and dynamic preferences of the POI.

\subsection{Local Differential Privacy (LDP)}
\label{local differential privacy}

LDP is a rigorous mathematical definition of privacy \cite{duchi2013local}, which is used to preserve the location privacy of users in our study. Previously, DP \cite{dwork2006calibrating} was employed in a centralized setting, which assumed the presence of a trusted data curator. In the centralized setting, each user submits their original data to the trusted data curator, and the curator randomizes the aggregated results to guarantee the privacy of the involved users. However, in the real world, we cannot guarantee that the data curator is always trusted. Hence, we adopted a local version of DP, which can preserve the privacy of users under an untrusted data curator.

LDP requires the following setting. There exists an untrusted server and $m$ users. A user $i$ holds a data item $x_{i}$ in his/her mobile device and the server wants to know the aggregated result of $x_{1}, x_{2}, \cdots, x_{m}$. To guarantee the privacy, each user $i$ is allowed to perturb $x_{i}$ to obtain a noisy version of $x_{i}$, say $x_{i}'$. Then, the server receives $x_{i}'$ instead of $x_{i}$, and calculates the aggregation result of $x_{1}', x_{2}', \cdots, x_{m}'$. While perturbing $x_{i}$, LDP requires a high probability that the server cannot infer the original value $x_{i}$ from the perturbed value $x_{i}'$. The probability is decided by a parameter $\varepsilon$, which controls the level of privacy guarantee. Specifically, LDP is defined as follows.

\begin{definition}{\textbf{$\varepsilon$-LDP}}
\label{def1}
A randomized mechanism $\mathcal{A}$ satisfies $\varepsilon$-LDP for any two input values $x_{1}, x_{2} \in Domain(\mathcal{A})$ and any possible output value $x'$ of $\mathcal{A}$, if the following condition is satisfied

\begin{equation*}
Pr[\mathcal{A}(x_{1})=x'] \leq e^{\varepsilon}Pr[\mathcal{A}(x_{2})=x'].
\end{equation*}

\end{definition}

Moreover, as LDP is also a variant of DP, LDP satisfies the composition theorem \cite{mcsherry2009privacy}. We introduce the sequential composition theorem that we utilized in our work as follows.

\begin{theorem}{\textbf{Sequential Composition}}
\label{theorem1}
Assume a randomized algorithm $\mathcal{F}$ consists of $k$ LDP mechanisms ($\mathcal{A}_{1},\cdots,\mathcal{A}_{k}$), where each satisfies ($\varepsilon_{1},\cdots,\varepsilon_{k}$)-LDP, respectively. Then, $\mathcal{F}$ guarantees $\sum_{i=1}^{k}{\varepsilon_{i}}$-LDP.
\end{theorem}

Because of the composition theorem, $\varepsilon$ is often referred to as privacy budget. Specifically, to guarantee $\varepsilon$-LDP, each LDP mechanism should use a part of $\varepsilon$, and the sum should be no more than $\varepsilon$.


\section{Problem definition}
\label{problem definition}

\subsection{Successive Point-of-Interest Recommendation}
\label{successive point-of-interest recommendation}

We first define the notion of \textit{POI}, \textit{check-in} and \textit{check-in history} as follows.

\begin{definition}{\textbf{POI}} A POI is defined as a specific location point uniquely identified with identifier $p_{i}$.
\label{def2}
\end{definition}

\begin{definition}{\textbf{Check-in}} A check-in is defined as a tuple $(u, p_{i}^{t})$, if user $u$ visited POI $p_{i}$ at timestamp $t$.

\label{def3}
\end{definition}

\begin{definition}{\textbf{Check-in history}} A check-in history of user $u$ over $T$ timestamps is defined as a sequence of check-ins represented as $CH_{u}=\{ (u, p_{x}^{1}), \cdots, (u, p_{y}^{T}) \}$.

\label{def4}
\end{definition}

Based on Definitions \ref{def2}, \ref{def3}, and \ref{def4}, we can define the successive POI recommendation problem as follows. Given a user $u$ and his/her check-in history $CH_{u}=\{ (u, p_{x}^{1}), \cdots, (u, p_{y}^{T}) \}$, the objective of the recommendation system is to recommend a set of POIs that the user $u$ is likely to visit at timestamp $T+1$.

\subsection{Challenges}
\label{challenges}

Our problem setting is challenging. In particular, our objective is to build a recommendation system under an untrusted server while ensuring the privacy of the check-in data. Moreover, when it comes to handling implicit feedback, existing privacy-preserving MF-based recommendation systems have a problem in interpreting non-negative values (in our case, visit counts) into useful preference values. The details of our challenges are as given below.

\begin{enumerate}
  \item \textit{What information can be shared for CD successive POI recommendation and how can it be shared?}  Existing studies on privacy-preserving recommendation systems \cite{hua2015differentially, zhang2018probabilistic, shin2018privacy} are suitable for SD scenarios with explicit feedback data. Considering that check-in actions are implicit feedback, we should model proper preference values from check-in data and integrate the preference values into MF.
  \item \textit{How can the check-in data be collected in a privacy-preserving manner?} A previous study \cite{de2013unique} showed that only four location points are sufficient to identify individuals. Therefore, people are hesitant to submit their check-in data, if the recommendation server is untrusted.
  \item \textit{How can the recommendation system be updated privately?} To accurately estimate the preference on the POIs, we should optimize the objective function of the proposed model. However, we cannot use the widely used DP-SGD technique \cite{abadi2016deep} that is designed for the centralized setting because an untrusted server can exploit the local gradients to reconstruct the input data of the model \cite{shin2018privacy}.
  \item \textit{How can the next POI candidates be received without disclosing the current location?} Although the collection of check-in data and the model updating is done in a privacy-preserving manner, the framework is not completely privacy-aware if the users who wish to receive services submit their exact location.
\end{enumerate}

\section{Proposed Method}
\label{proposed method}

\subsection{Overview of our framework}
\label{overview of our framework}

In this section, we introduce our SPIREL framework illustrated in Figure \ref{fig1}. In this section, we will use the terms ``user'' and ``participant'' interchangeably. In our proposed framework, we assumed that $m$ participants exist, each having their individual check-in histories, and an untrusted recommendation server. Note that the check-in data is a type of implicit feedback and it provides only positive samples (i.e., visited POIs). Thus, if we only use the visit counts to model the interactions between users and POIs, the user-POI matrix is extremely sparse as the users typically visit only a small number of POIs.

\Figure[ht]()[scale=0.56]{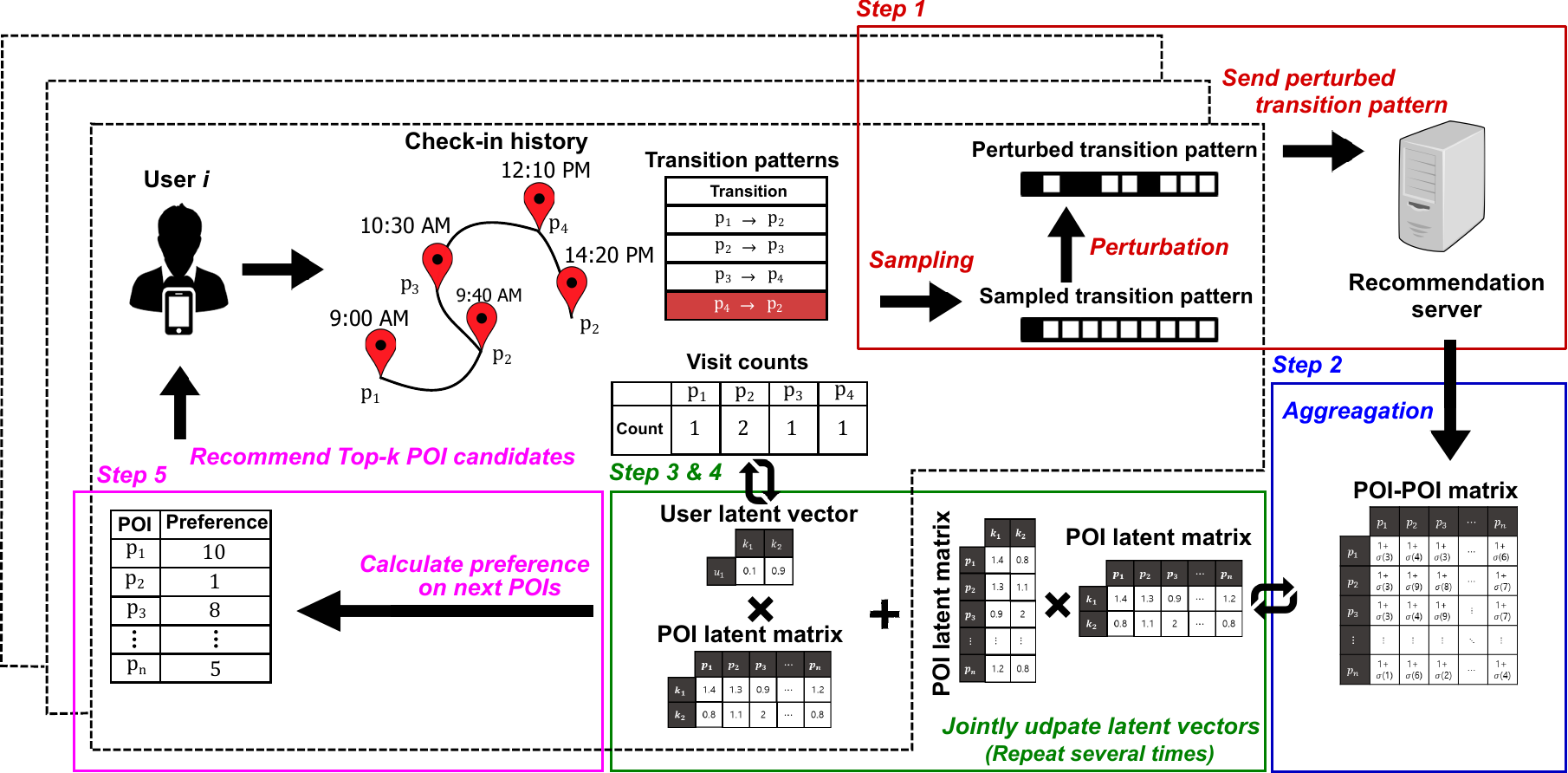}
{Overview of SPIREL framework \label{fig1}}

To infer the interest for the next POIs based on the current location, we used the notion of a first-order Markov chain to model the sequential transition pattern between two consecutive POIs. For example, a user in Figure \ref{fig1} has four transition patterns ($p_{1} \rightarrow p_{2}$, $p_{2} \rightarrow p_{3}$, $p_{3} \rightarrow p_{4}$, $p_{4} \rightarrow p_{2}$).

\Figure[ht]()[width=3in]{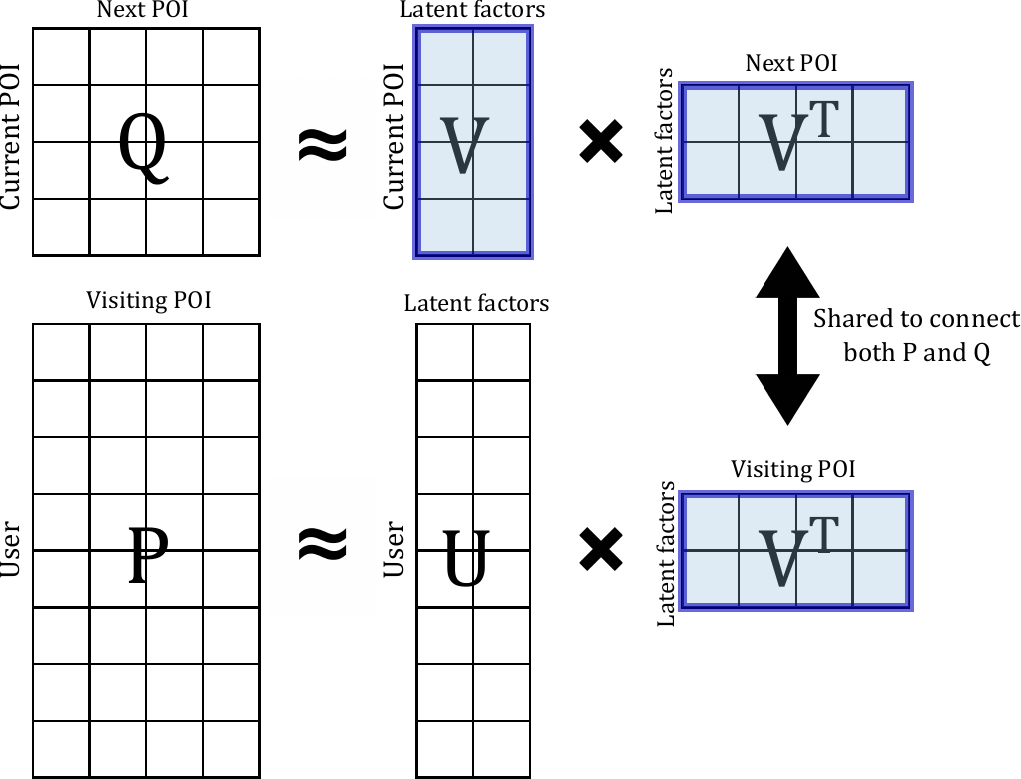}
{Transfer learning in SPIREL framework\label{fig2}}

Figure \ref{fig2} graphically illustrates the transfer learning approach in SPIREL. Here, the user-POI matrix $P$ represents the POI preference of users based on the visit counts, and the POI-POI matrix $Q$ is learned from the collected POI transition patterns. Then, the POI latent matrix $V$ is shared and used to connect both the user-POI matrix $P$ and POI-POI matrix $Q$. Thus, the visit counts and the transition patterns are integrated into the learning process of MF. Ultimately, our primary idea is that the knowledge learned from the transition patterns (matrix $Q$) can help increase the next POI recommendation accuracy in the target domain (matrix $P$). Accordingly, the objective function of SPIREL should be revised to learn $P$ and $Q$ simultaneously compared to Equation \ref{eq1}. We describe our novel objective function as follows.

\begin{equation} 
\label{eq5}
\begin{aligned}
\mathcal{L}_{SPIREL} = ||P - UV^{\intercal}||^{2} & + ||Q - VV^{\intercal}||^{2} \\
& + \lambda (||U||^{2} + ||V||^{2})
\end{aligned}
\end{equation}

Here, we explain the process flow of our proposed framework. 

\begin{itemize}
  \item \textbf{Step 1.} Each of the $m$ participants independently samples a single transition pattern between two consecutive POIs. Then, the participant perturbs the sampled transition pattern in their local device and submits the perturbed transition pattern to the server.
  \item \textbf{Step 2.} After receiving the perturbed transition patterns from all participants, the server builds a POI-POI matrix that approximately reflects the global transition trend.
  \item \textbf{Step 3.} The server performs the learning process, which minimizes the objective function (Equation \ref{eq5}). First, the users update their profile vector using ALS. Then, each user perturbs his/her local gradients of POIs and submits them to the server.
  \item \textbf{Step 4.} The server aggregates the perturbed gradients received from the users and updates the POI latent matrix using SGD.
  \item \textbf{Step 5.} The server sends the learned POI latent matrix to each user. Then, the users can calculate the preference score of the next POIs by themselves.
\end{itemize}

In the following section, we will explain each of these steps in detail.

\subsection{Transition Pattern Aggregation under LDP}
\label{transition pattern aggregation under LDP}

\Figure[ht]()[width=6.7in]{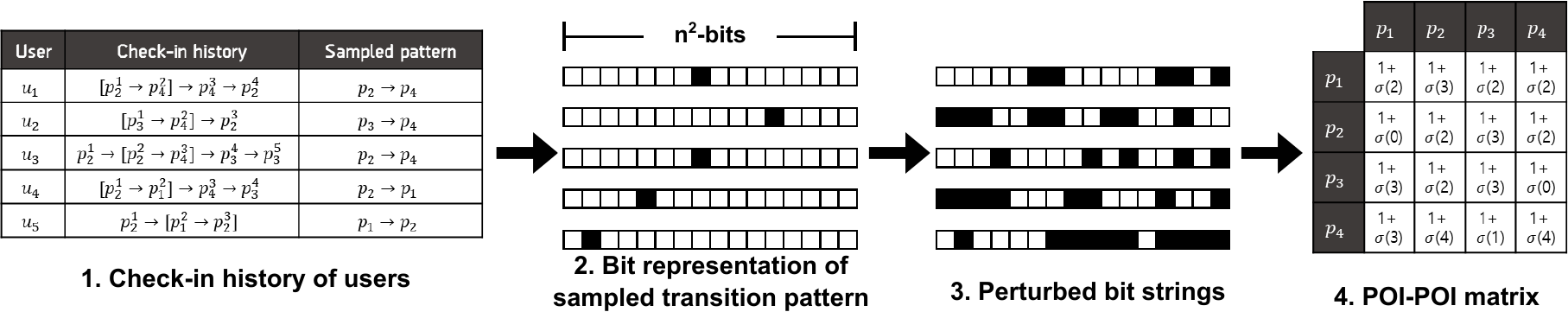}
{Example of point-of-interest (POI) transition pattern aggregation under local differential privacy (LDP)\label{fig3}}

\subsubsection{Participants (Step 1)}
\label{transition pattern perturbation:Participants}
Suppose there are $n$ POIs and the check-in history consists of $T$ check-ins. Then, the possible number of check-in histories is $n^{T}$. A naive solution would be to directly collect the frequency of each possible check-in history. However, directly computing frequencies over an enormous domain for all possible check-in histories is impractical, even if $n$ and $t$ are very small.

Our idea is to sample one transition pattern from the check-in history and estimate the frequency of sampled transition patterns under LDP to capture the global transition trend. Figure \ref{fig3} shows an example of the transition pattern aggregation process, where each bit represents one specific transition pattern. We assume that all users share the same POI domain, and each user selects a single transition pattern from his/her check-in history. In this way, we can prevent certain users who have a long sequence of check-ins from dominating the global transition trend. Moreover, it makes the server harder to re-identify the users, as information about only two POIs is reported.

\begin{algorithm}[!t]
\caption{Optimized randomized response (ORR)}
\label{alg1}
    \begin{algorithmic}[1]
        \Require privacy budget $\varepsilon$, binary value $b[i]\in \{0,1 \}$
        \Ensure perturbed binary value $\hat{b}[i]$
        
        \State Set $p = \frac{1}{2}$, $q = \frac{1}{e^{\varepsilon} + 1}$
        
        \If {$b[i]=1$} $\hat{b}[i] \thicksim Bernoulli(p)$
        \Else {} $\hat{b}[i] \thicksim Bernoulli(q)$
        \EndIf

        \State return $\hat{b}[i]$

	\end{algorithmic}
\end{algorithm}

Randomized response (RR) \cite{warner1965randomized} is a basic mechanism for guaranteeing LDP, where a user submits a ``yes'' or ``no'' answer according to the result of a coin-flipping process. The purpose of RR is to estimate the actual number of ``yes'' users. We used the optimized RR (ORR) scheme \cite{wang2017locally}, which minimizes the estimation error. We present the detailed process of ORR in Algorithm \ref{alg1}.

\begin{algorithm}[!t]
\caption{Perturbing transition patterns using ORR (client side)}
\label{alg2}
    \begin{algorithmic}[1]
        \Require privacy budget $\varepsilon$, the number of POIs $n$
        \Ensure perturbed bit string $\hat{b}$
        
        \State User $u$ randomly samples a transition pattern $p_{x}^{t} \rightarrow p_{y}^{t+1}$ from check-in history
		    
        \State User $u$ initializes a length $n^2$ bit string $b$
        \For{$i=1$ to $n^2$}
            \If {$p_{x} \rightarrow p_{y}$ corresponds to the $i$th position of $b$} $b[i]=1$
            \Else {} $b[i]=0$
            \EndIf
            
            \State $\hat{b}[i]=ORR(\varepsilon, b[i])$ \Comment{Algorithm \ref{alg1}}

        \EndFor
        \State Send $\hat{b}$ to the server

	\end{algorithmic}
\end{algorithm}

Algorithm \ref{alg2} shows the detailed transition pattern perturbation process. In our framework, each user first generates a binary bit string of length $n^{2}$, where each bit represents whether the sampled transition pattern corresponds to the specific POI-POI combination. As directly sending the original bit value can cause a location privacy breach, each user perturbs the bit value using the ORR mechanism (Algorithm \ref{alg1}). Here, ORR is invoked $n^{2}$ times to collect the support of each transition pattern from all users.

When reporting the transition patterns, the communication cost $O(n^2)$ could be a problem if the domain size of POI is too large. If the communication cost is critical, we could reduce this cost by instead adopting one-bit randomized response methods, such as Hadamard randomized response technique \cite{apple2014learning, nguyen2016collecting, acharya2019hadamard} following the analysis in \cite{kim2018learning}.

\subsubsection{Recommendation server (Step 2)}
\label{transition pattern perturbation:recommendation server}

\begin{algorithm}[!t]

\caption{Aggregating perturbed transition patterns (server side)}
\label{alg3}
    \begin{algorithmic}[1]
        \Require $m$ perturbed bit strings $\hat{b}$, the number of POIs $n$
	    \Ensure a POI-POI matrix $Q$
	    
	    \State Initialize a $n \times n$ matrix $Q$ whose elements are all zero
	   
		\For{$i=1$ to $n$}
			\For{$j=1$ to $n$}
			    \For{$k=1$ to $m$}
                    \State $Q[i][j] = Q[i][j] + \hat{b}_{k}[j + n*i]$
                \EndFor
                
                \State $Q[i][j] = \frac{Q[i][j] - mq}{p-q}$ \Comment{estimate the frequency of each transition pattern}
                
                \State $Q[i][j] = 1 + \sigma(Q[i][j])$ \Comment{calculate confidence score of each transition pattern}
		    \EndFor
	    \EndFor
    
		\State return $Q$
	\end{algorithmic}
\end{algorithm}

The server aggregates the perturbed bit strings and estimates the frequency of each transition pattern. In Algorithm \ref{alg3}, we present the procedure for aggregating the perturbed transition patterns to build a POI-POI matrix $Q$, which is used as the auxiliary information for the learning process. After receiving the perturbed bit strings from $m$ users, the server adds each bit value to the corresponding element of $Q$. As the users implement ORR to perturb their transition patterns, the server can estimate the frequency of each transition pattern by calculating $\frac{Q[i][j] - mq}{p-q}$, where $p$ and $q$ are predefined probability in Algorithm \ref{alg1}. 

Note that we let each user sample and perturb one transition pattern from his/her check-in history as discussed in Section \ref{transition pattern perturbation:Participants}. Thus, it is challenging how to interpret the estimated frequency and map the fake positive feedback to user preference values. Here, we follow an assumption that positive feedback (no matter real or fake) simply indicates 'more preferable' than non-observed feedback \cite{rendle2012bpr}.

Randomized response methods essentially give more weight to the probability that the truthful answer is delivered as it is. However, we cannot directly use the estimated frequency as preference scores because of the following two reasons. First, the estimated frequency obtained through the randomized response methods can have negative values. If the true count is close to zero, the noisy count can be unbiasedly estimated and thus can be a negative value. In this case, we cannot assume that the negative value indicates that all the participants dislike the corresponding transition pattern. Secondly, the estimated frequency grows linearly with the number of participants. Thus, we should carefully transfer the estimated frequency into another preference scores. 

A widely used approach is to use confidence scores as preferences, which represents how much confidence increases with the number of positive feedback. For example, \cite{hu2008collaborative} suggested a plausible choice for the confidence score as $c_{ui}=1+\alpha r_{ui}$. Here, $r_{ui}$ indicates the number of actions of user $u$ on item $i$. The rate of increase in confidence is controlled by $\alpha$. In our proposed method, we used the sigmoid function and set the ($i,j$)-th element of matrix $Q$ to $1 + \sigma(Q[i][j])$. The intuitive reasons for using the sigmoid function are as follows. First, the sigmoid function takes all real numbers as the input; thus, it can handle the dynamic range of estimated frequency, including negative value. Second, its output is monotonically increasing and is bounded in the range $[0,1]$, which can handle any number of participants.

\Figure[ht]()[width=6.5in]{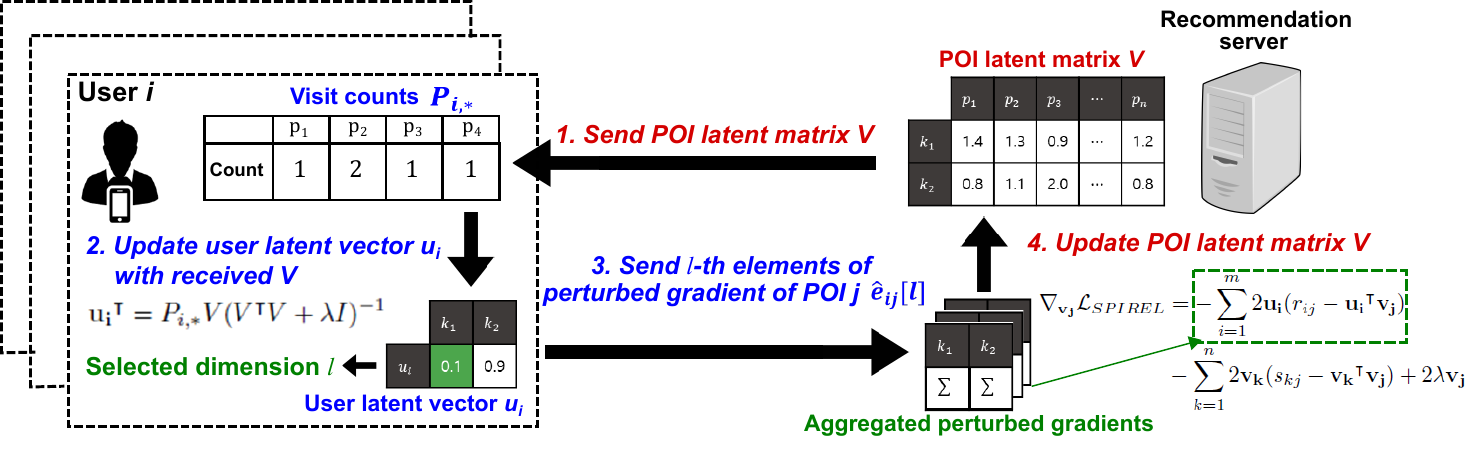}
{Learning process of SPIREL by gradient perturbation under LDP\label{fig4}}

\subsection{Privacy-Preserving Learning under LDP}
\label{gradient perturbation}

In this section, we propose a locally private solution to optimize our objective function. After building a POI-POI matrix, the next step is to factorize two matrices to identify the latent factors of users and POIs. Note that our objective function (Equation \ref{eq5}) aims to factorize the user-POI and POI-POI matrices simultaneously. Accordingly, we are required to derive the update rules for Equation \ref{eq5}.

In Section \ref{matrix factorization}, we introduced two techniques to minimize the quadratic objective function. SPIREL adopts both SGD and ALS to optimize Equation \ref{eq5}. The previous work \cite{shin2018privacy} used SGD only to update both user and item profile vectors. However, SGD typically requires more iterations compared to ALS, and its performance is very sensitive to the choice of the learning rate. There is a limit on the maximum iterations in the privacy-preserving recommendation system since the gradient computation at each iteration consumes the privacy budget, which can sometimes achieve an unsatisfactory model.

We first update the user profile vectors using ALS because it converges faster than SGD. Another reason is that each user can independently update their profile vector, which is more suitable for distributed learning setting \cite{tan2016faster}. Then, we can obtain the ALS update rule for the profile vector of user $i$ as follows.

\begin{equation} 
\label{eq6}
\begin{split}
\frac{\partial \mathcal{L}_{SPIREL}}{\partial \mathbf{u_{i}}} &= -2\sum_{j}  (r_{ij}-\mathbf{u_{i}}^{\intercal}\mathbf{v_{j}})\mathbf{v_{j}}^{\intercal} + 2\lambda \mathbf{u_{i}}^{\intercal} \\
\mathbf{u_{i}}^{\intercal} &= P_{i,*}V(V^{\intercal}V + \lambda I)^{-1}
\end{split}
\end{equation}

As shown in Equation \ref{eq6}, the user profile update rule is the same as in Equation \ref{eq4}. As we assumed the POI profile vectors as a constant, only the user-POI matrix term remains. Here, the server can pre-compute $V(V^{\intercal}V + \lambda I)^{-1}$ and send the term to the users. In this way, the users do not have to forward their visit counts ($P_{i,*}$) to the server to update their profile vector. Next, we list the ALS update rule for the POI profile vector as follows.

\begin{equation} 
\label{eq7}
\begin{split}
\frac{\partial \mathcal{L}_{SPIREL}}{\partial \mathbf{v_{j}}} &= -2\sum_{i}  (r_{ij}-\mathbf{v_{j}}^{\intercal}\mathbf{u_{i}})\mathbf{u_{i}}^{\intercal} \\
& -2 \sum_{k}(s_{kj}-\mathbf{v_{j}}^{\intercal}\mathbf{v_{k}})\mathbf{v_{k}}^{\intercal}
+ 2\lambda \mathbf{v_{j}}^{\intercal} \\
0 &= -(P_{*,j}^{\intercal} - \mathbf{v_{j}}^{\intercal}U^{\intercal})U  \\
& -(Q_{*,j}^{\intercal} - \mathbf{v_{j}}^{\intercal}V^{\intercal})V + \lambda \mathbf{v_{j}}^{\intercal} \\
P_{*,j}^{\intercal}U + Q_{*,j}^{\intercal}V &= \mathbf{v_{j}}^{\intercal}(U^{\intercal}U + V^{\intercal}V + \lambda I) \\
\mathbf{v_{j}}^{\intercal} &= (P_{*,j}^{\intercal}U + Q_{*,j}^{\intercal}V)(U^{\intercal}U + V^{\intercal}V + \lambda I)^{-1}
\end{split}
\end{equation}

Here, $P_{*,j}$ indicates a column vector of matrix $P$ with index $j$ and $s_{kj}$ indicates the $(k,j)$-th element of matrix $Q$. The server requires a user latent matrix $U$ to update the POI profile vector, which can reveal the interactions between users and POIs by multiplying the POI latent matrix $V$ with user latent matrix $U$. One option is to let users add noise to their profile vector in each iteration of training. However, this method will significantly affect both recommendation accuracy and convergence of learning, since the noise added to the profile vectors can lead to a domino effect of adding noise to the outputs of MF algorithm.

For circumventing the above issue, we instead perturbed the gradients and applied SGD to update the POI profile vectors. In this way, we can bound the noise to affect only the learning process of MF. Specifically, we let each user submit the perturbed gradients of Equation \ref{eq5}, as displayed in Figure \ref{fig4}. Then, the recommendation server aggregates the perturbed gradients and updates the POI profile vectors. Finally, we can rewrite the gradient of the POI profile vector as follows.

\begin{equation} 
\label{eq8}
\begin{split}
\nabla_{\mathbf{v_{j}}}\mathcal{L}_{SPIREL} & = 
-\sum_{i=1}^{m} 2\mathbf{u_{i}}(r_{ij}-\mathbf{u_{i}}^{\intercal}\mathbf{v_{j}})\\
& - \sum_{k=1}^{n} 2\mathbf{v_{k}}(s_{kj}-\mathbf{v_{k}}^{\intercal}\mathbf{v_{j}})
+ 2\lambda\mathbf{v_{j}}
\end{split}
\end{equation}

\subsubsection{Participants (Step 3)}
\label{gradient perturbation:participants}

\begin{algorithm}[!t]
\caption{Perturbing gradients (client side)}
\label{alg4}
    \begin{algorithmic}[1]
        \Require privacy budget $\varepsilon$, gradient value of POI $j$ $e_{ij}$, the number of POIs $n$, profile vector size $d$, a randomly selected dimension $l$ of $e_{ij}$
	    \Ensure perturbed gradient $\hat{e_{ij}}[l]$
	    
	    \State Project $e_{ij}[l]$ into the range [-1, 1]
    
        \State User $i$ draws a value $x$ from Bernoulli distribution such that Pr$[x=1]=\frac{e_{ij}[l] \cdot (e^{\varepsilon}-1) + (e^{\varepsilon}+1)}{2e^{\varepsilon}+2}$
        
        \If{$x = 1$}
            \State $\hat{e_{ij}}[l]=nd\frac{e^{\varepsilon}+1}{e^{\varepsilon}-1}$
        \Else{}
            \State $\hat{e_{ij}}[l]=-nd\frac{e^{\varepsilon}+1}{e^{\varepsilon}-1}$
        \EndIf
        
        \State return $\hat{e_{ij}}[l]$
	\end{algorithmic}
\end{algorithm}

Equation \ref{eq8} predominantly consists of two terms: $\mathbf{u_{i}}(r_{ij}-\mathbf{u_{i}}^{\intercal}\mathbf{v_{j}})$ and $\mathbf{v_{k}}(s_{kj}-\mathbf{v_{k}}^{\intercal}\mathbf{v_{j}})$. The recommendation server can calculate the term $\mathbf{v_{k}}(s_{kj}-\mathbf{v_{k}}^{\intercal}\mathbf{v_{j}})$ by itself because the value $s_{kj}$ is the confidence scores, as discussed in Section \ref{transition pattern aggregation under LDP}. The other term $\mathbf{u_{i}}(r_{ij}-\mathbf{u_{i}}^{\intercal}\mathbf{v_{j}})$ has a prediction error of the visit counts $r_{ij}$. In our framework, the participants perturb the term $\mathbf{u_{i}}(r_{ij}-\mathbf{u_{i}}^{\intercal}\mathbf{v_{j}})$ to prevent the recommendation server from learning whether a user $i$ visits POI $j$. 

In our framework, we applied the widely used RR method proposed by Nguyen et al. \cite{nguyen2016collecting} to perturb the user gradients, which was also utilized in \cite{shin2018privacy}. It should be noted that SPIREL is not confined to a specific LDP mechanism; thus, any LDP mechanisms, such as \cite{wang2019collecting} and \cite{duchi2014privacy}, can be used in our proposed framework. In the experiments, we presented the results using a more accurate LDP mechanism \cite{wang2019collecting} and demonstrated that the error induced by noisy gradients has a limited effect on the performance.

Algorithm \ref{alg4} shows the procedure of the gradient perturbation process. Each participant first chooses a random value of $l$ and $j$, thereby sending only a single dimension of the perturbed gradient for a single randomly chosen POI. Because the gradient values are not bounded, we impose a bound on the raw gradients before perturbation by manually clipping the gradients at each iteration \cite{abadi2016deep}. As the LDP mechanism we use assumes that the input value lies in the range $[-1,1]$, we project the gradient $e_{ij}$ in $[-1,1]$. Then, the user generates a noisy gradient $\hat{e_{ij}}$ following the probability distribution of LDP mechanism and submits the perturbed gradient to the server. Note that $\hat{e_{ij}}$ needs to be scaled by a factor of $n$ and $d$ because of sampling $l$ and $j$ respectively in Algorithm \ref{alg5}.

\subsubsection{Recommendation server (Step 4)}
\label{gradient perturbation:recommendation server}

\begin{algorithm}[!t]
\caption{Update of point-of-interest latent matrix (server side)}
\label{alg5}
    \begin{algorithmic}[1]
        \Require the number of POIs $n$, the number of iterations $I$, profile vector size $d$
	    \Ensure updated POI latent matrix $V$
        
        \State Divide the users into $I$ groups \{${g_{1}, g_{2}, \cdots, g_{I}}$\}
        \For{$i=1$ to $I$}
            \State Initialize $\nabla_{\mathbf{V}}\mathcal{L}_{SPIREL} \in \{0\}^{n \times d}$
            \For{each user $u$ in group $g_{i}$}
                    \State Selects a random value $j$ from $\{1,2,\cdots,n\}$
                    \State Selects a random value $l$ from $\{1,2,\cdots,d\}$
                    
                    \State Gets perturbed gradient $\hat{e_{uj}}[l]$ \Comment{Algorithm \ref{alg4}}
                    \State$\nabla_{\mathbf{V}}\mathcal{L}_{SPIREL}$ = $\nabla_{\mathbf{V}}\mathcal{L}_{SPIREL} + \hat{e_{uj}}[l]$
            \EndFor
            \State Compute other terms in Equation \ref{eq8} and add to $\nabla_{\mathbf{V}}\mathcal{L}_{SPIREL}$
            \State Update $V$ with $\nabla_{\mathbf{V}}\mathcal{L}_{SPIREL}$ using Adam optimizer
        \EndFor
        
        \State return $V$
	\end{algorithmic}
\end{algorithm}

The server is responsible for aggregating the perturbed gradients from the participants and updating the POI profile vectors. As each user submits a noisy gradient of a randomly chosen POI through Algorithm \ref{alg4}, the server cannot learn about the details of the POIs visited by the user. However, the server can still estimate the sum of perturbed gradients and thus updates the model. We further applied two strategies to update our models efficiently.

\textbf{(1) Learning with user group} Generally, the non-private recommendation systems will stop the learning process when the objective function converges. However, under a privacy-preserving setting, the server continues the learning process during a predefined number of iterations because the participants are supposed to submit perturbed gradients at each iteration. For example, \cite{shin2018privacy} sets a fixed number of iterations $i$. At each iteration, each user submits a perturbed gradient using $\frac{\varepsilon}{i}$ privacy budget, which guarantees the entire learning process satisfies $\varepsilon$-LDP.

When answering multiple questions, Wang et al. \cite{wang2017locally} proved that partitioning users into groups and assigning one question for each group is better than splitting the privacy budget in terms of accuracy of the aggregated result under LDP. Inspired by this idea, we divided the users randomly into $i$ groups and asked a single user group to submit their perturbed gradients using the entire privacy budget. In this way, each user only participates in one iteration of the learning process. This approach is also in line with the mini-batch gradient descent concept \cite{li2014efficient}, where the model is updated with a subset of training examples during one iteration.

\textbf{(2) Stabilizing iterative learning} After obtaining a gradient value, the next step is to determine a learning rate that decides how fast the model learns. If the learning rate is set too high, the training path of objective function frequently oscillates and may never converge. On the other hand, if the learning rate is too low, the model will take too many iterations to converge. In our framework, determining the learning rate is more challenging due to the inherent noisy gradients.

Recently, variants of gradient descent methods that have the power of adaptive learning rates were widely applied to train deep neural networks, such as Adagrad \cite{duchi2011adaptive}, RMSprop \cite{hinton2012lecture}, and Adam \cite{kingma2014adam}. Here, we used Adam optimizer that maintains a weighted moving average of the gradients (Equation \ref{eq9}). Averaging over gradients makes the whole learning process stable and robust to initial learning rate selection.

\begin{equation}
\label{eq9}
\begin{split}
x &= \beta_{1}x + (1-\beta_{1})\nabla_{\mathbf{v_{j}}}\mathcal{L}_{SPIREL} \\
y &= \beta_{2}y + (1-\beta_{2})(\nabla_{\mathbf{v_{j}}}\mathcal{L}_{SPIREL})^{2}
\end{split}
\end{equation}

Here, as $x$ and $y$ are typically initialized to zero, the Adam optimizer performs bias correction as follows.

\begin{equation}
\label{eq10}
\hat{x} = \frac{x}{1-\beta_{1}},  \hat{y} = \frac{y}{1-\beta_{2}} 
\end{equation}

Finally, the update rule is provided by the following equation.

\begin{equation}
\label{eq11}    
\mathbf{v_{j}} = \mathbf{v_{j}} - \frac{\gamma}{\sqrt{\hat{y}}+\epsilon}\hat{x}
\end{equation}

Here, $\beta_{1}$ and $\beta_{2}$ control the decay rates of the moving averages. Further, $\gamma$ indicates the learning rate and $\epsilon$ is used to avoid division by zero.

\subsection{Ranking (Step 5)}
\label{ranking}

Using the learned profile vector of users and POIs, SPIREL can recommend the next POI candidates while considering the current location of users. Let us assume a user $i$ is in POI $j$. Then, the preference of next POI $k$ of user $i$ can be calculated as follows.

\begin{equation}
\label{eq12}
    pref_{i}^{k} = \mathbf{u_{i}}^{\intercal}\mathbf{v_{k}} + \mathbf{v_{j}}^{\intercal}\mathbf{v_{k}}
\end{equation}

Equation \ref{eq12} consists of two preferences: the first term, $\mathbf{u_{i}}^{\intercal}\mathbf{v_{k}}$, indicates the personal preference for the next POI $k$ and the second term, $\mathbf{v_{j}}^{\intercal}\mathbf{v_{k}}$, represents the POI transition preference of POI $j$ to POI $k$. By calculating the sum of the two preferences, we can recommend the top-k next POIs to users. It should be noted that because the POI profile vectors are publicly known, users are not required to submit their current location to the server. Instead, the recommendation server sends all the POI profile vectors to users, and the users can compute the preferences for all next POIs within their local device.

\subsection{Complexity Analysis}
\label{complexity analysis} 

In this subsection, we analyze the computational complexity of the participants and the server in the SPIREL framework, respectively.

\begin{itemize}
    \item \textbf{Participants.} Each participant first perturbs the $n^2$-length bit string. So the time complexity of perturbing bit string is $O(n^2)$. After sending the bit string, each user updates their profile according to Equation \ref{eq6}. Note that the server can precompute the $n \times d$ matrix $V(V^{\intercal}V + \lambda I)^{-1}$ in advance. Thus, each user simply needs to multiply $n$-length vector $P_{i,*}$ with $n \times d$ matrix, which requires $O(nd)$ time. Finally, each user calculates the $d$-length gradient vector in the first term of Equation \ref{eq8} and perturbs one bit of the gradient vector, which can be finished in $O(d^2)$ time. In summary, the total time complexity of one user is $O(n^2)$.
    \item \textbf{Server.} The server first aggregates the $n^2$-length bit string from $m$ users and constructs the $n \times n$ POI-POI matrix. The time complexity for building the POI-POI matrix is $O(mn^2)$. Next, the server precomputes the term $V(V^{\intercal}V + \lambda I)^{-1}$ used for the ALS. Even though matrix conversion of $d \times d$ matrix $(V^{\intercal}V + \lambda I)^{-1}$ requires $O(d^3)$ time, $n$ is typically larger than $d$. Thus, we can assume that the server needs $O(nd^2)$ time to compute the entire term. Finally, the server updates the POI latent matrix according to Equation \ref{eq8}. Assume that the maximum number of iterations is $i$. For each iteration, the server computes the second term of Equation \ref{eq8}, which requires $O(nd^2)$ time. Then, the total time complexity for updating the POI latent matrix is $O(ind^2)$. Since $m$ is much larger than $i$ or $d$, we can conclude that the total time complexity of the server is $O(mn^2)$.
\end{itemize}

\subsection{Privacy Analysis}
\label{privacy analysis}

In this subsection, we analyze the privacy guarantee of each step in SPIREL. Our framework mainly consists of two perturbation processes, as explained in Sections \ref{transition pattern aggregation under LDP} and \ref{gradient perturbation}. According to Theorem \ref{theorem1}, to guarantee $\varepsilon$-LDP, the users have to split their privacy budget for each process. In our work, we let users use $\varepsilon_{1}$ to perturb their transition patterns and $\varepsilon_{2}$ to perturb a gradient value of selected POI; thus the sum of $\varepsilon_{1}$ and $\varepsilon_{2}$ equals to $\varepsilon$.

\begin{enumerate}
    \item \textbf{Step 1.} Each participant prepares a $n^{2}$-length bit string, where only a single bit is set according to the sampled transition pattern. Other bits are zero. Each participant invokes ORR mechanism $n^{2}$ times to report supports of each transition pattern to the server. As each bit is relevant to the independent transition pattern, the participant can use the entire budget to perturb each bit. Thus, the process of perturbing such binary vectors satisfies $\varepsilon_{1}$-LDP as analyzed in \cite{wang2017locally}.
    \item \textbf{Step 2.} The server aggregates the perturbed bit strings received from all participants. Then, the server estimates the supports of transition patterns. Based on the estimated supports, the server computes the preference score for each transition pattern. Because of DP's post-processing immunity, no further privacy leaks occur in this step.
    \item \textbf{Step 3.} First, the server transmits the information needed to update the user profile vector through ALS using POI latent matrix. Then, each participant can locally update their profile vector without any transmission. Second, each user submits a perturbed gradient value for a randomly selected POI. Note that each user is involved in a disjoint user group, and only a single user group is assigned to one iteration of the learning procedure. The participants clip their one sampled gradient value to the range [-1, 1] and spend the entire $\varepsilon_{2}$ budget for perturbation. We can exploit any LDP mechanism in this step, and the privacy guarantee depends on the LDP mechanism used. In our work, we utilize the LDP mechanism proposed by Nguyen et al. \cite{nguyen2016collecting}, and it ensures $\varepsilon_{2}$-LDP.
    \item \textbf{Step 4.} For each iteration of the learning procedure, the server receives the perturbed gradients from one dedicated user group and updates the POI profile vectors. Note that the participant submits a perturbed gradient for a randomly selected POI and its value is either $\pm nd\frac{e^{\varepsilon}+1}{e^{\varepsilon}-1}$. Thus, the server cannot infer which POIs are visited by the participants or how much they visited those POIs. Therefore, this step does not involve any privacy disclosure.
    \item \textbf{Step 5.} The server publishes the updated profile vectors of all POIs to the participants. Then, each participant can compute the preference score of the next POI without disclosing their current position. Since the server does not know the profile vectors of participants, there is no privacy leakage.
\end{enumerate}


\section{Evaluation}
\label{evaluation}

In this section, we evaluate the SPIREL method in various settings. All experiments were performed on a server with Intel i9-9900X CPU, 128GB of memory, and GeForce RTX 2080 GPU. In all experiments, we averaged the results over 50 runs.

\subsection{Experimental Setup}
\label{Experimental setup}

\begin{table}[!t]
\caption{Statistics of four datasets}
\label{table2}
\centering
\begin{tabular}{|c|c|c|c|c|}
\hline
Dataset & Users & POIs & Check-ins & Sparsity \\
\hline
\hline
Gowalla & 9,617 & 585 & 95,956 & 98.29\%\\
\hline
Taxi & 267,739 & 526 & 5,354,780 & 96.20\%\\
\hline
Yelp & 97,141 & 1,075 & 691,744 & 99.34\%\\
\hline
Foursquare & 32,158 & 2,409 & 423,675 & 99.45\%\\
\hline
\end{tabular}
\end{table}

\subsubsection{Datasets}
\label{datasets}

We used four public datasets: \textit{Gowalla}\footnote{https://snap.stanford.edu/data/loc-gowalla.html}, \textit{Taxi}\footnote{https://data.cityofchicago.org/Transportation/Taxi-Trips/wrvz-psew/data}, \textit{Yelp}\footnote{https://www.yelp.com/dataset} and \textit{Foursquare} \cite{yin2016adapting}. For each of the above datasets, we discard users and POIs with fewer than 10 associated check-ins. Table \ref{table2} shows the statistics for all datasets after cleaning. We excluded the last check-in for the evaluation and trained the model with the remaining check-ins for each check-in history. Here, data sparsity is based on $1-\frac{Check-ins}{Users \cdot POIs}$.

\subsubsection{Methods}
\label{methods}

SPIREL is the first work to integrate the relationship between POIs with that of users and POIs while guaranteeing the LDP. There is no existing privacy-preserving recommendation method that can directly compare with SPIREL. Accordingly, our ultimate goal is to demonstrate the benefits of combining the POI-POI relationship in MF and the privacy-preserving learning process of SPIREL compared to the following five methods.

\begin{enumerate}
    \item \textbf{SD approach} This approach is a single-domain MF-based recommendation system that only considers the visit counts to predict the preference of the next POIs.
    \item \textbf{SD approach with private learning (SD-PL) \cite{shin2018privacy}} This is a private version of SD, which assumes that users hold visit counts in their local devices. SD-PL optimizes the same objective function of SD except that the users follow the LDP protocol of \cite{shin2018privacy}, i.e., gradients are sent to the server after perturbation.
    \item \textbf{CD approach} This is a non-private version of our framework, which assumes that the server receives the raw check-in history of users. Thus, the CD approach directly uses original visit counts and all transition patterns (not sampled, not perturbed) between two POIs. 
    \item \textbf{SPIREL with piecewise mechanism (SPIREL-PM)} This is a variant of SPIREL that utilizes a piecewise mechanism \cite{wang2019collecting} in the stage of gradient perturbation. Note that the piecewise mechanism is known to obtain higher accuracy than the LDP mechanism \cite{nguyen2016collecting} used in SPIREL.
    \item \textbf{Item-based collaborative filtering (ICF) \cite{guo2019locally}} This approach predicts the next items based on the item similarity score. The item similarity score is derived from the estimated joint frequencies of co-occurrence between two items using the LDP mechanism.
\end{enumerate}

\subsubsection{Metrics}
\label{metrics}

We employed two evaluation metrics to evaluate the quality of recommendation: hit ratio (HR) and mean reciprocal rank (MRR). HR@$k$ measures the accuracy of the recommendation, which indicates the ratio of users whose latest POI is included in the given top-$k$ POI candidates recommended by the server. MRR is another widely used metric for evaluating the quality of recommendation. For a single recommendation result, the reciprocal rank is $\frac{1}{rank}$, where $rank$ is the position of the correct answer in the recommendation list. We evaluated HR@$k$ and MRR for $k$ in $\{3,5,7,10\}$, and the default value of $k$ was set to 5.

\subsubsection{Parameters}
\label{parameters}

We experimented with five values of privacy budget $\varepsilon$ in $\{0.2,0.4,0.8,1.6,3.2\}$, where the default value is 0.8. The regularization parameter $\lambda$ was set to $10^{-4}$. We set the size of the profile vector $d$ to 40 for all methods. The learning rate $\gamma$ was set to 0.01 for CD, SPIREL, and SPIREL-PM. In SD and SD-PL, $\gamma$ was set to 0.001. We also considered various values for the maximum number of iterations from 5 to 40 (by default 20) and different privacy budget allocation ratios (we divided the budget equally by default). The parameters related to the Adam optimizer were set to $\beta_{1}=0.9$, $\beta_{2}=0.999$, and $\epsilon=10^{-8}$ based on \cite{kingma2014adam} in CD and SPIREL variants. Finally, we used the same parameter setting for ICF in \cite{guo2019locally}.

\subsection{Experimental Results}
\label{experimental results}

\begin{figure*}[!t]
    \centering
    \null\hfill
    \subfloat[Gowalla]{
        \includegraphics[scale=0.10]{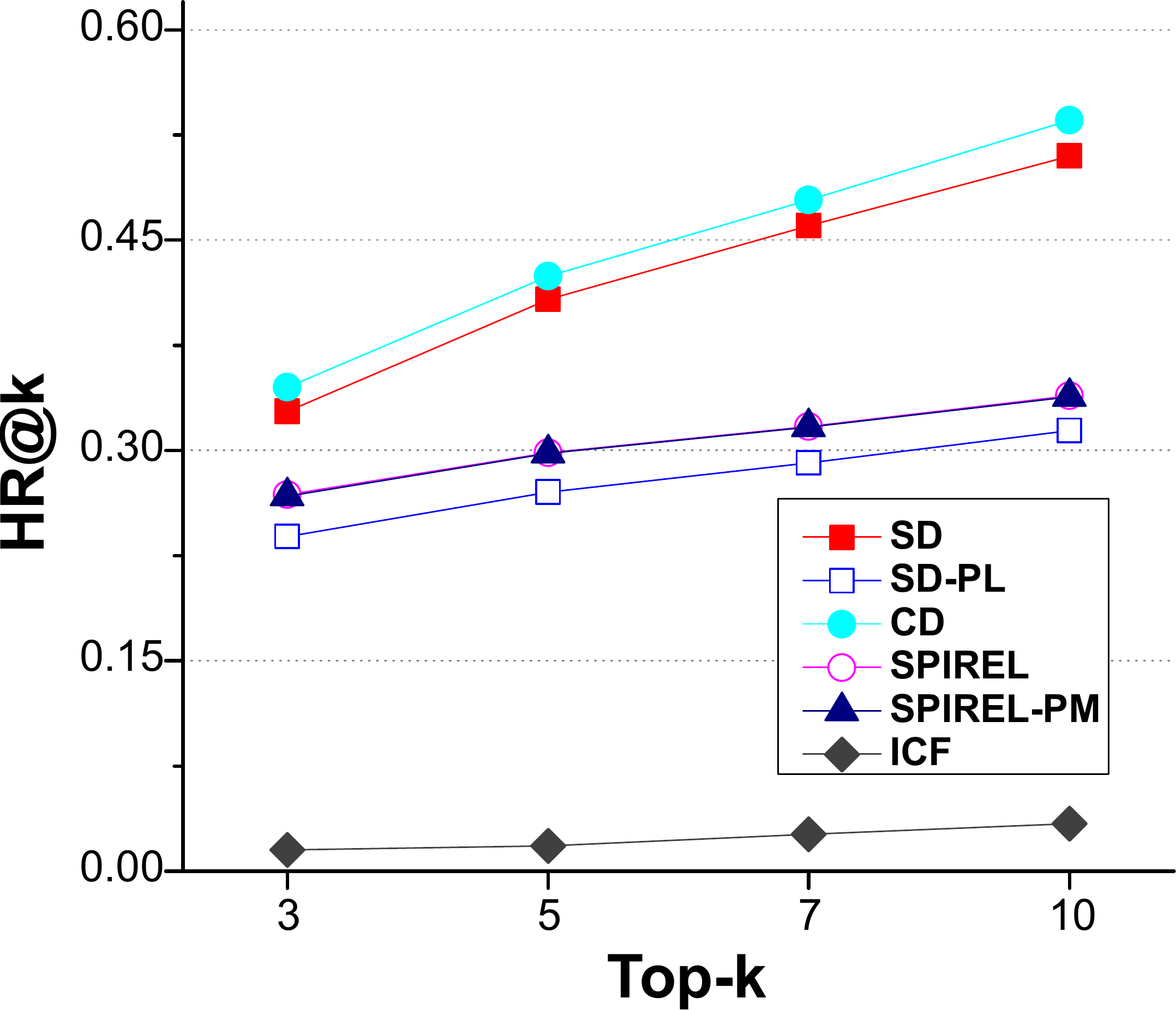}
        \label{fig5-1}
    }
    \hfill
    \subfloat[Taxi]{
        \includegraphics[scale=0.10]{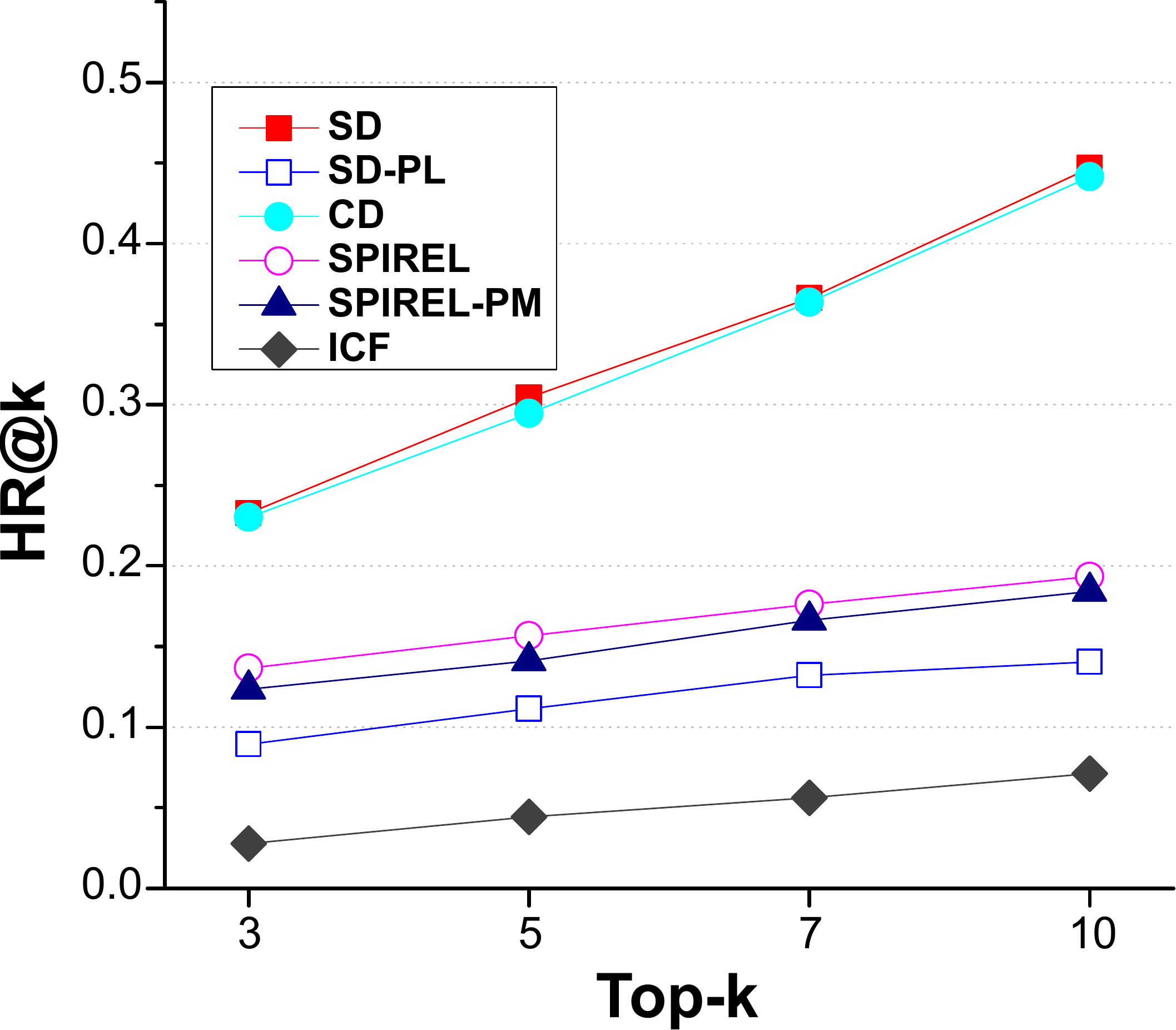}
        \label{fig5-2}
    }
    \hfill
    \subfloat[Yelp]{
        \includegraphics[scale=0.10]{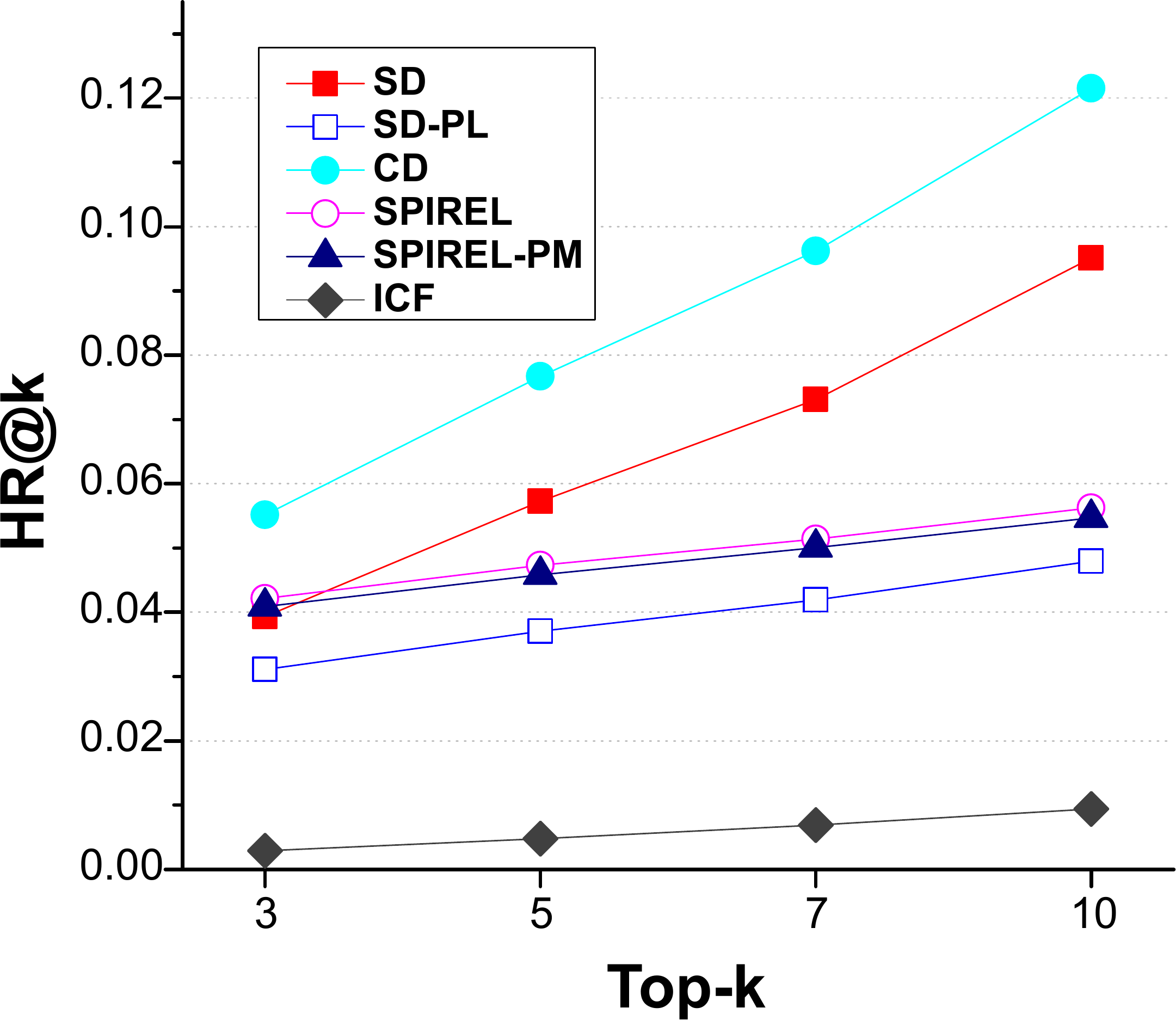}
        \label{fig5-3}
    }
    \hfill
    \subfloat[Foursquare]{
        \includegraphics[scale=0.10]{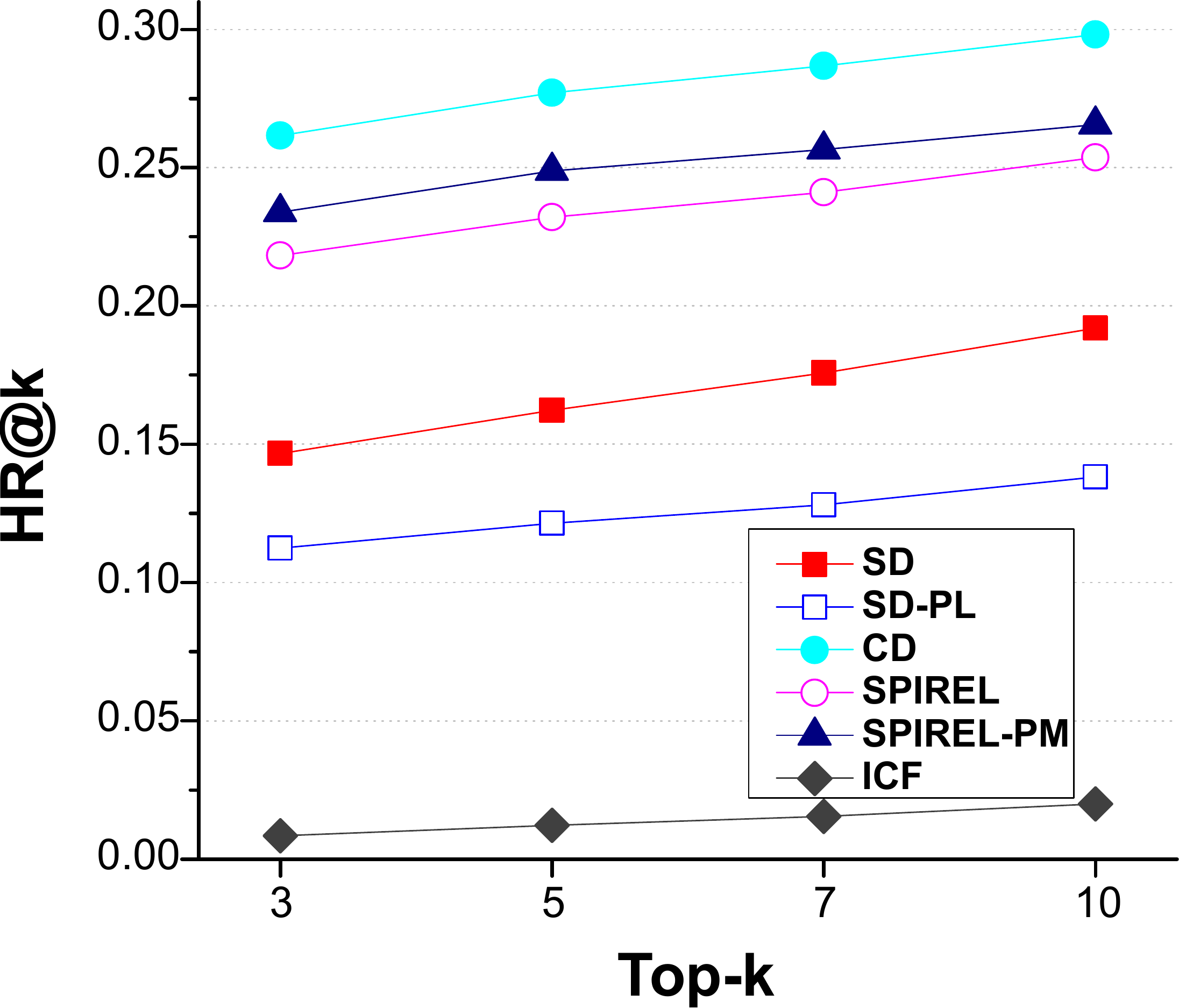}
        \label{fig5-4}
    }
    \null\hfill
    \caption{HR@k comparison with varying top-k recommendation sizes}
    \label{fig5}
\end{figure*}

\begin{figure*}[!t]
    \centering
    \null\hfill
    \subfloat[Gowalla]{
        \includegraphics[scale=0.10]{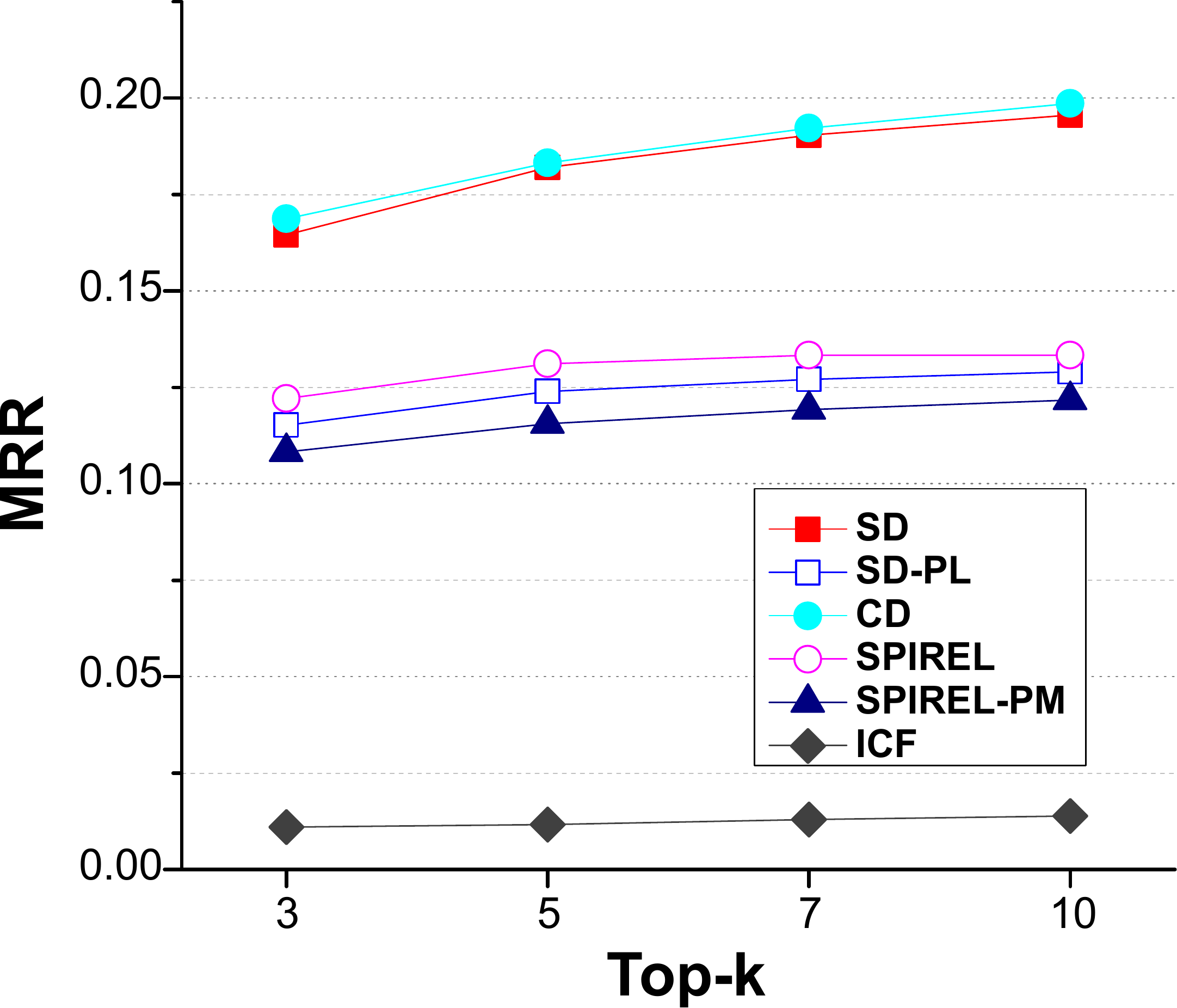}
        \label{fig6-1}
    }
    \hfill
    \subfloat[Taxi]{
        \includegraphics[scale=0.10]{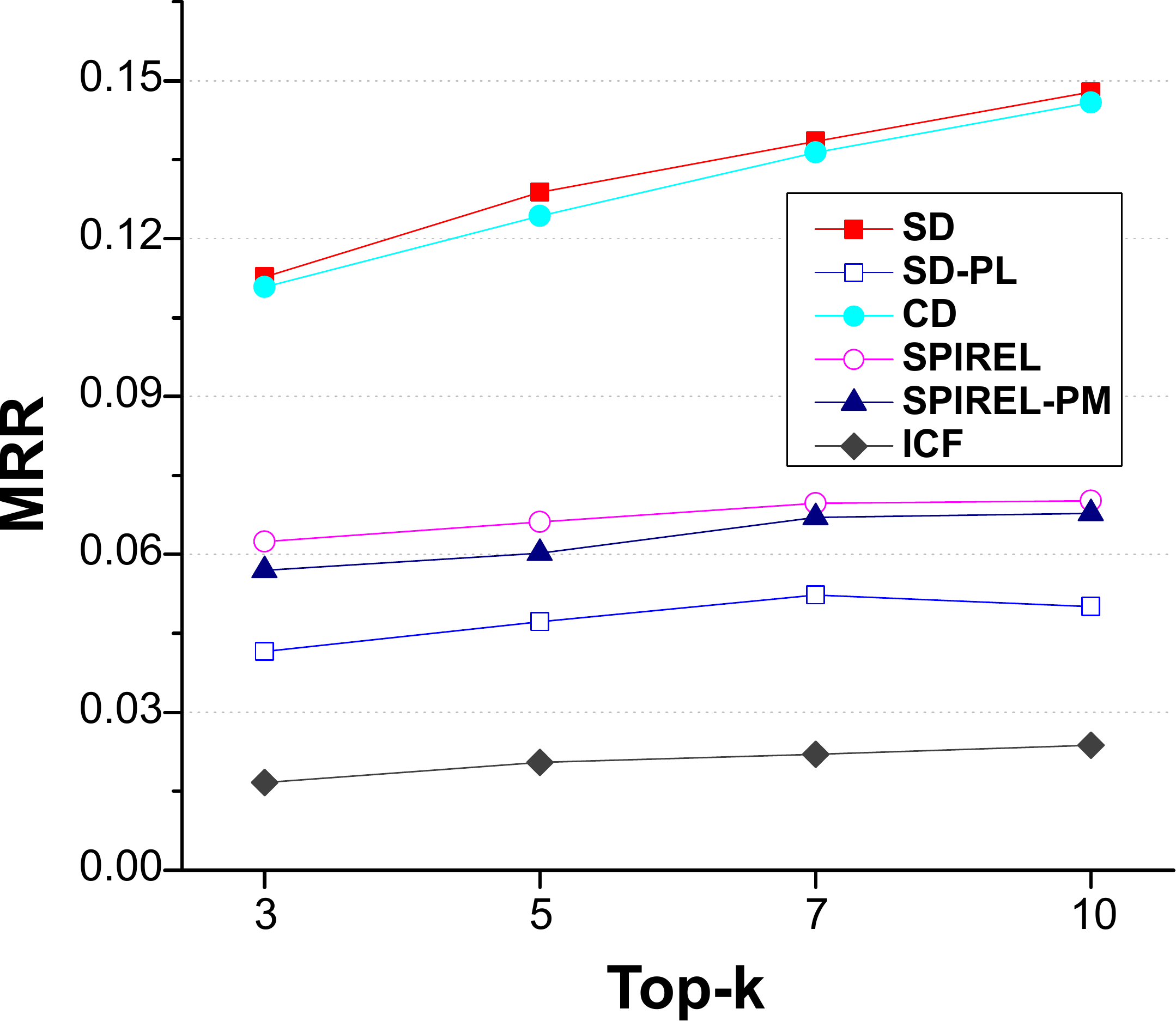}
        \label{fig6-2}
    }
    \hfill
    \subfloat[Yelp]{
        \includegraphics[scale=0.10]{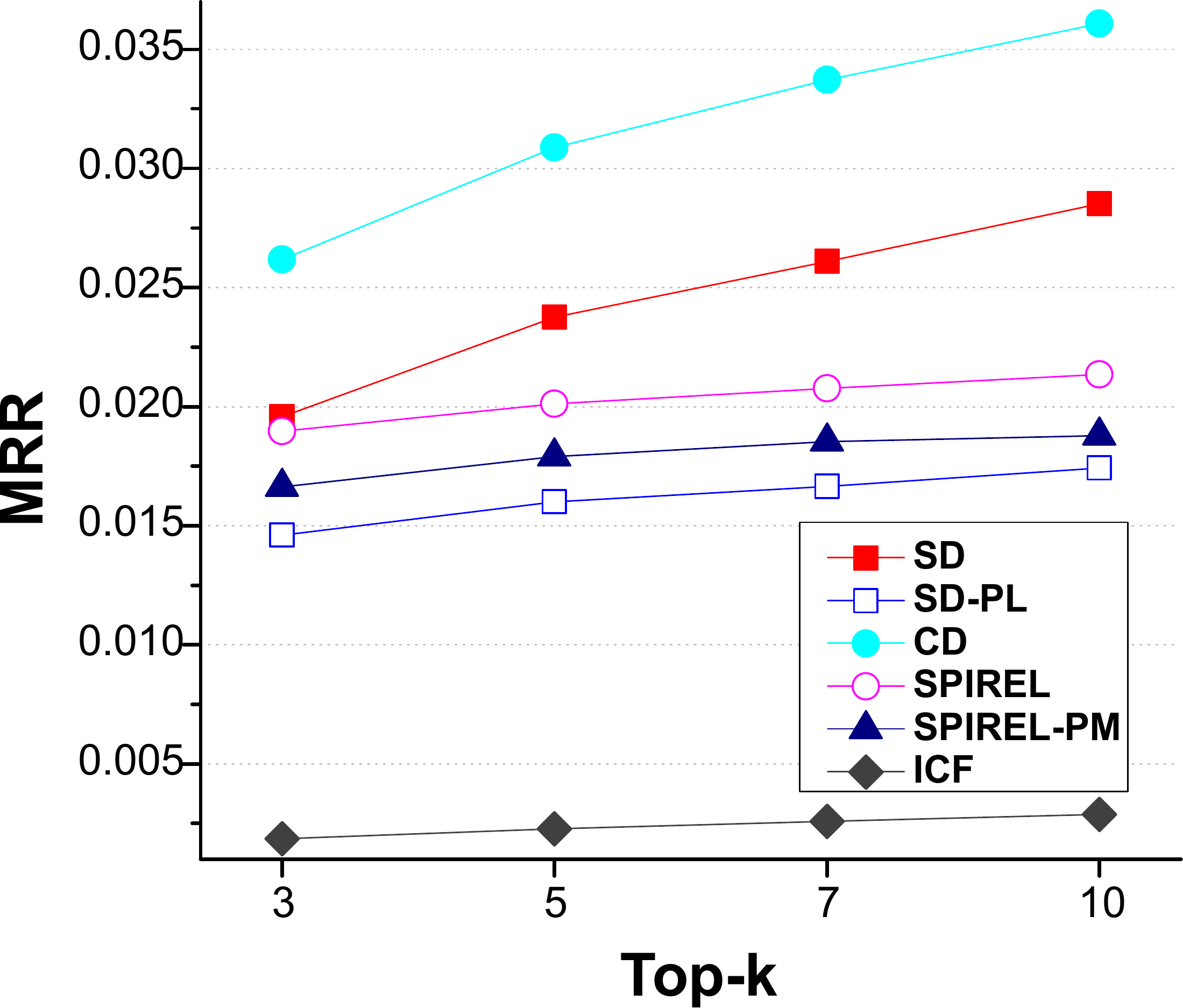}
        \label{fig6-3}
    }
    \hfill
    \subfloat[Foursquare]{
        \includegraphics[scale=0.10]{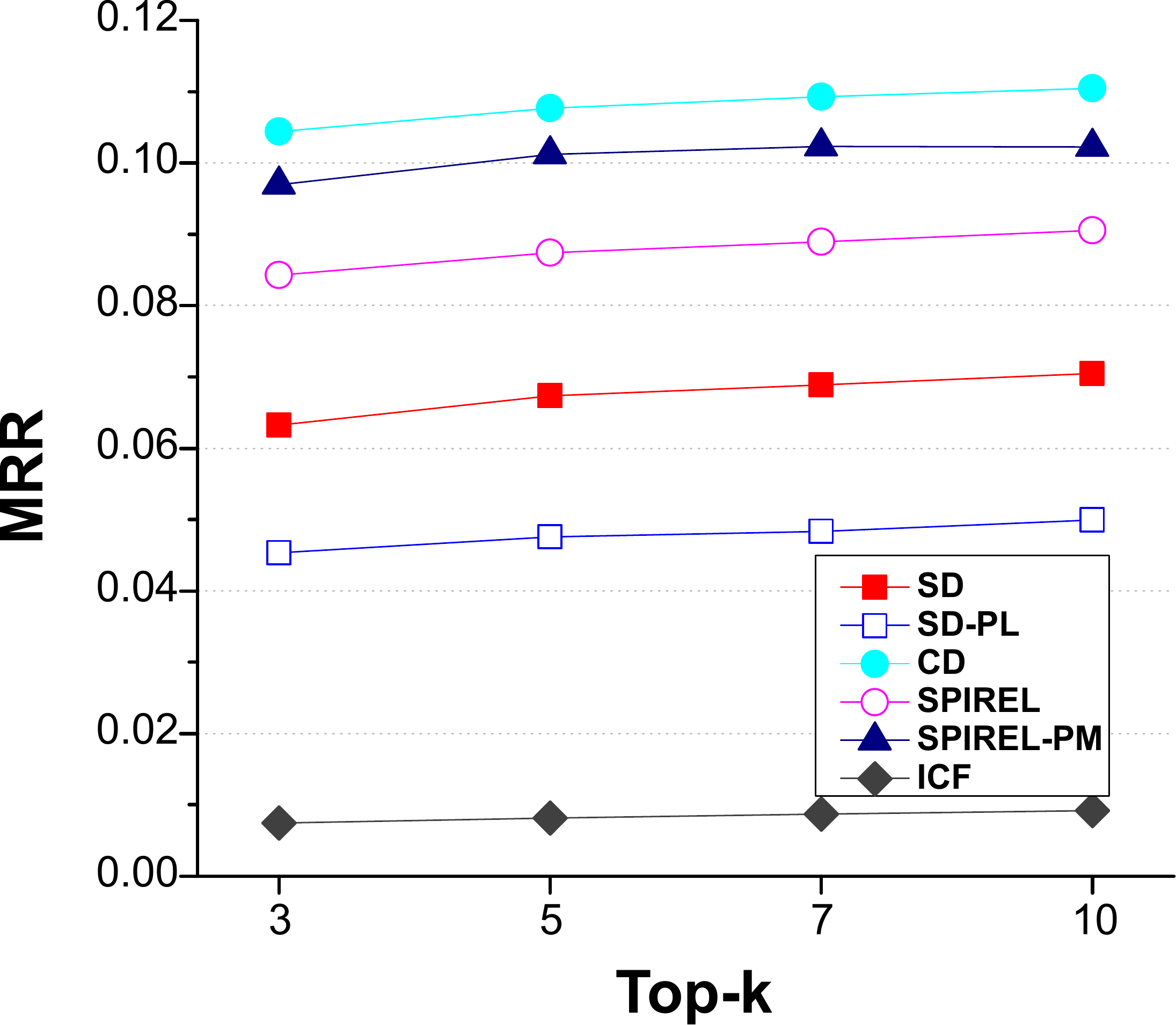}
        \label{fig6-4}
    }
    \null\hfill
    \caption{MRR comparison with varying top-k recommendation sizes}
    \label{fig6}
\end{figure*}

\subsubsection{Varying recommendation sizes}
\label{varying recommendation sizes}

Figs. \ref{fig5} and \ref{fig6} show the HR@$k$ and MRR with different recommendation list sizes, respectively. We first compared the non-private methods (SD and CD) to identify the effects of integrating the POI-POI relationship. Overall, we observed the CD approach outperforms on all datasets in HR@k and MRR, except for the \textit{Taxi} dataset. The reason is that the sparsity of \textit{Taxi} dataset is very low compared to other datasets, which allows the SD approach to predict next POIs sufficiently with only the user-POI relationship. Concretely, CD approach improves the SD by 25.82\% in HR@$k$ and 22.35\% in MRR on average.

Next, we compared SD-PL and SPIREL, which are the private version of SD and CD, respectively. We can observe both methods experience performance drops due to the perturbation process. Nevertheless, SPIREL always outperforms SD-PL, on average, 41.56\% and 38.84\% in terms of HR@$k$ and MRR, respectively. Another observation is that SPIREL even achieves better performance than SD in \textit{Foursquare} dataset, presumably due to the extreme sparsity of the user-POI domain. Compared to methods that rely on the user-POI relationship alone, the results show the advantages of integrating the confidence score of the noisy POI-POI relationship.

ICF consistently shows the worst performance among all methods. Note that ICF does not consider the users' preferences for POIs. Thus, it is beneficial to unify both the user-POI and POI-POI relationships for modeling users' preferences. Finally, SPIREL-PM achieves similar performance compared to SPIREL. This is due to the situation in which SPIREL should divide the privacy budget, and we will discuss this in detail in Section \ref{varying privacy budget allocation ratios}.

\begin{figure*}[!t]
    \centering
    \null\hfill
    \subfloat[Gowalla]{
        \includegraphics[scale=0.10]{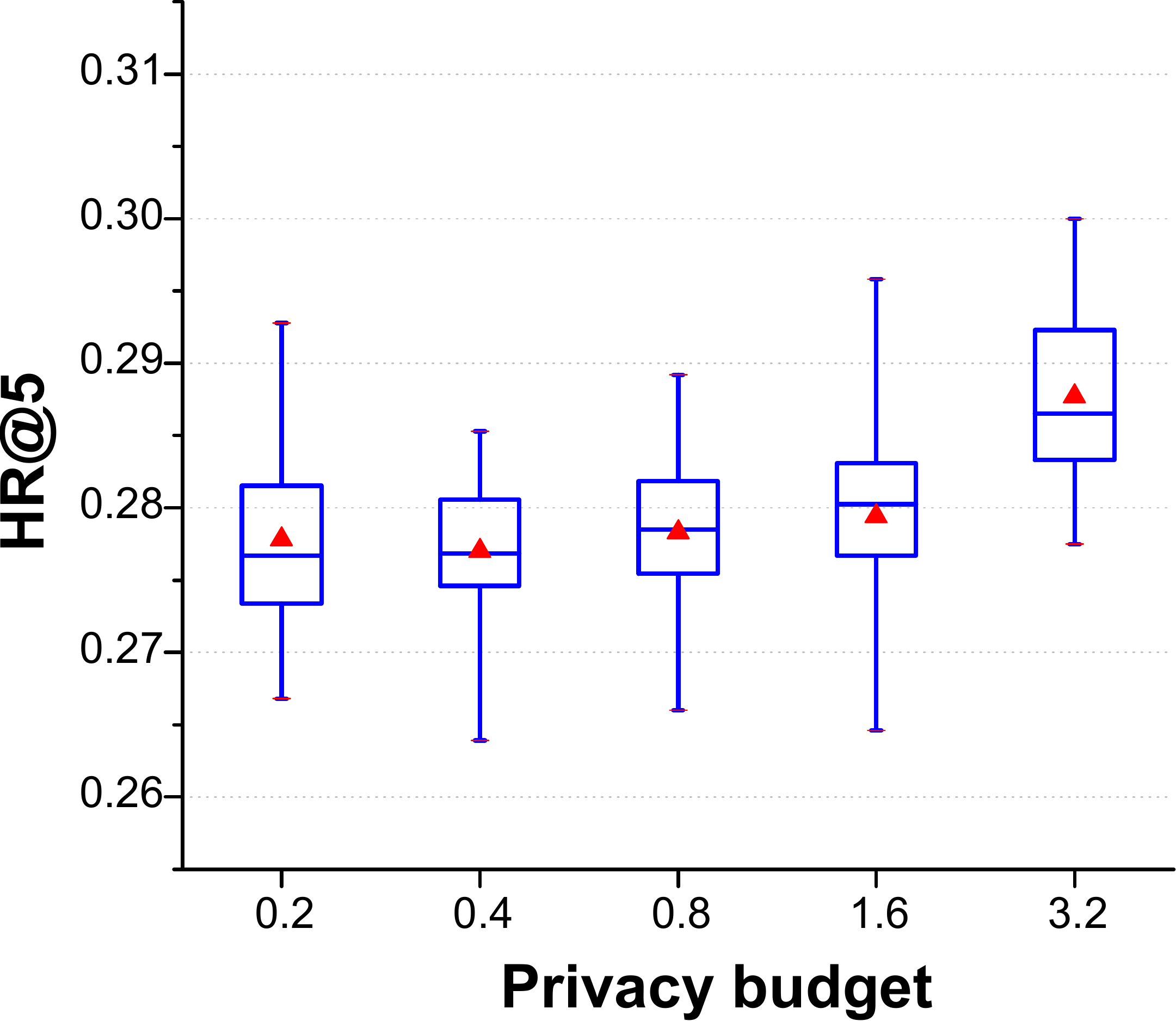}
        \label{fig7-1}
    }
    \hfill
    \subfloat[Taxi]{
        \includegraphics[scale=0.10]{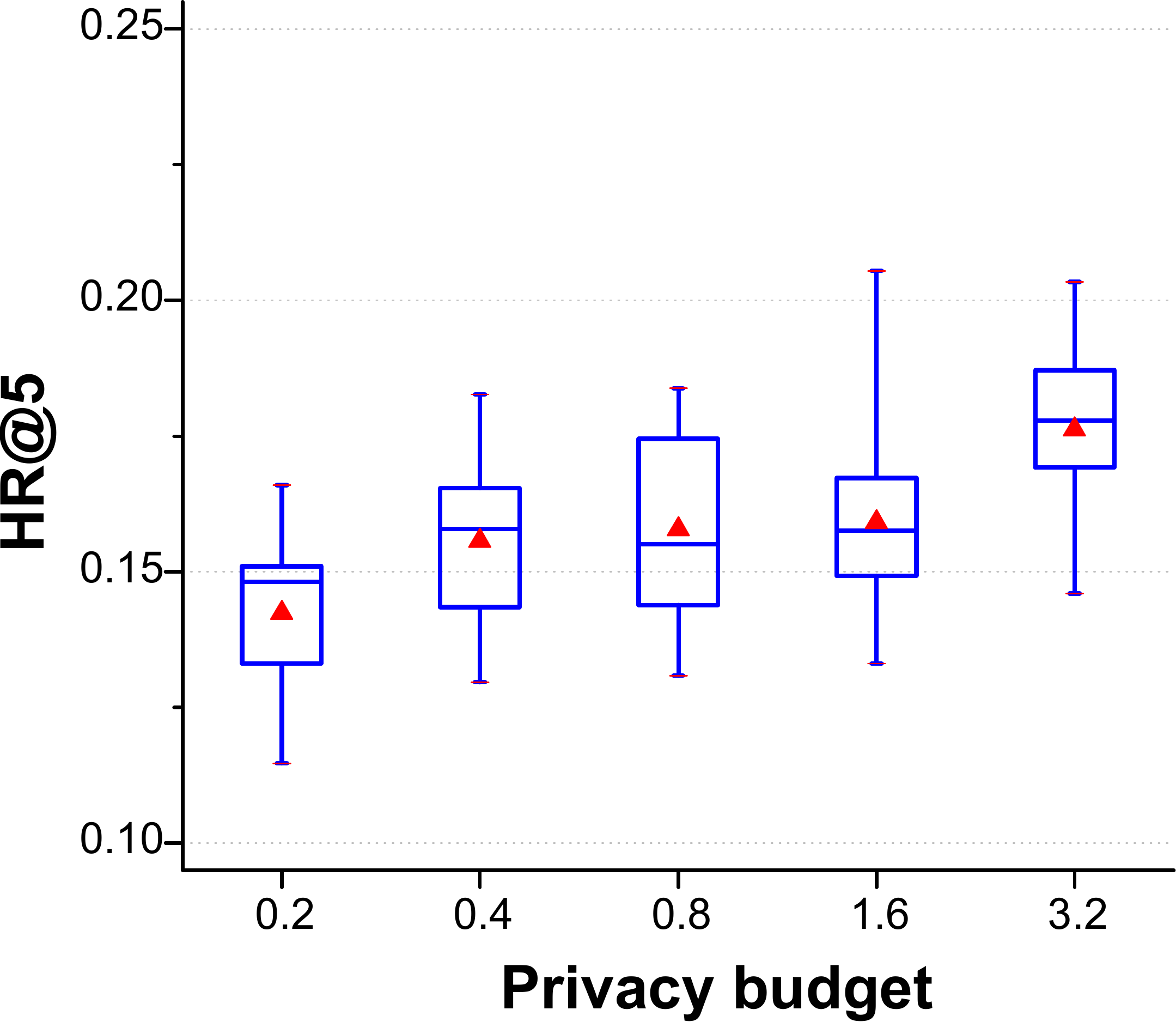}
        \label{fig7-2}
    }
    \hfill
    \subfloat[Yelp]{
        \includegraphics[scale=0.10]{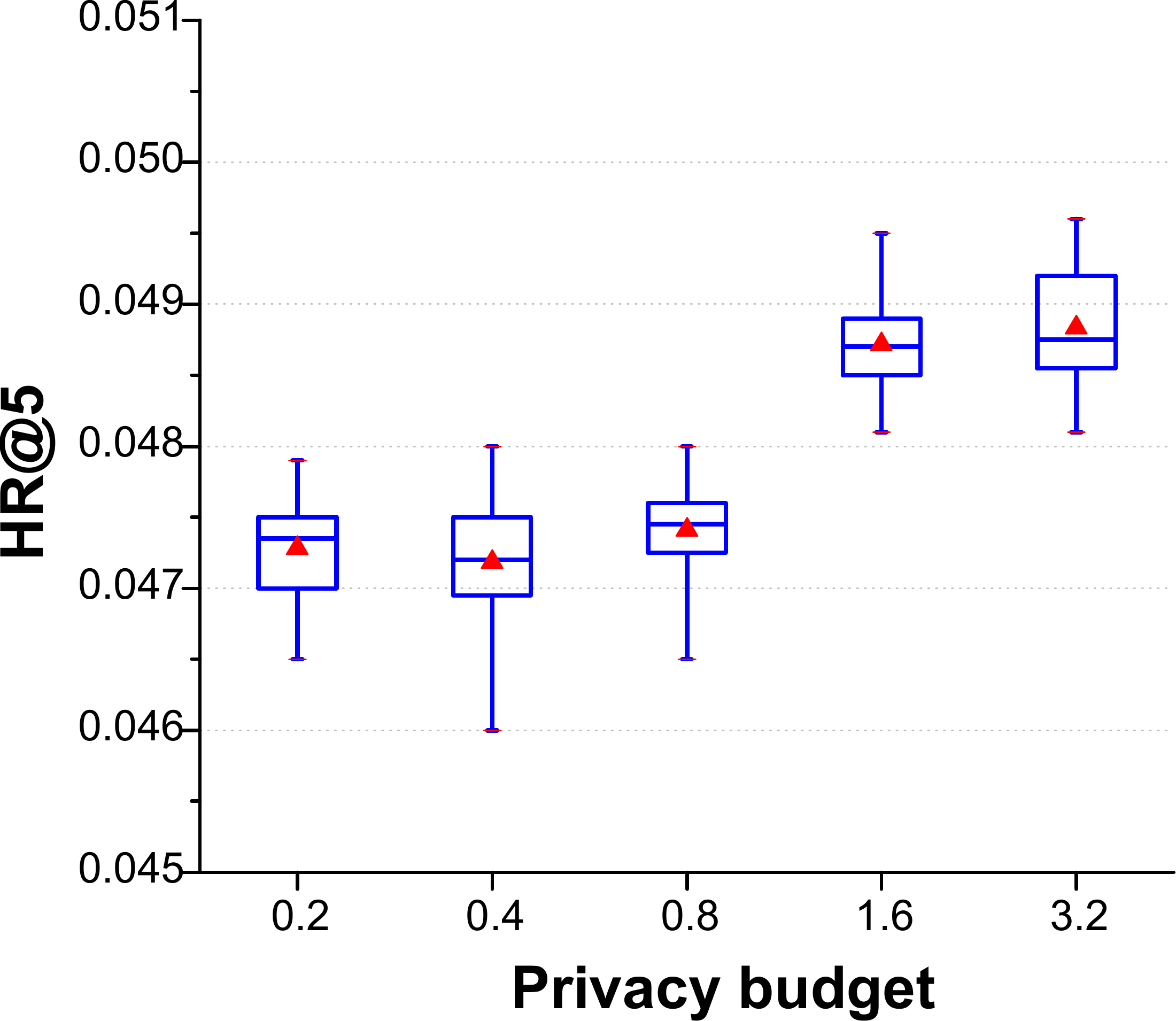}
        \label{fig7-3}
    }
    \hfill
    \subfloat[Foursquare]{
        \includegraphics[scale=0.10]{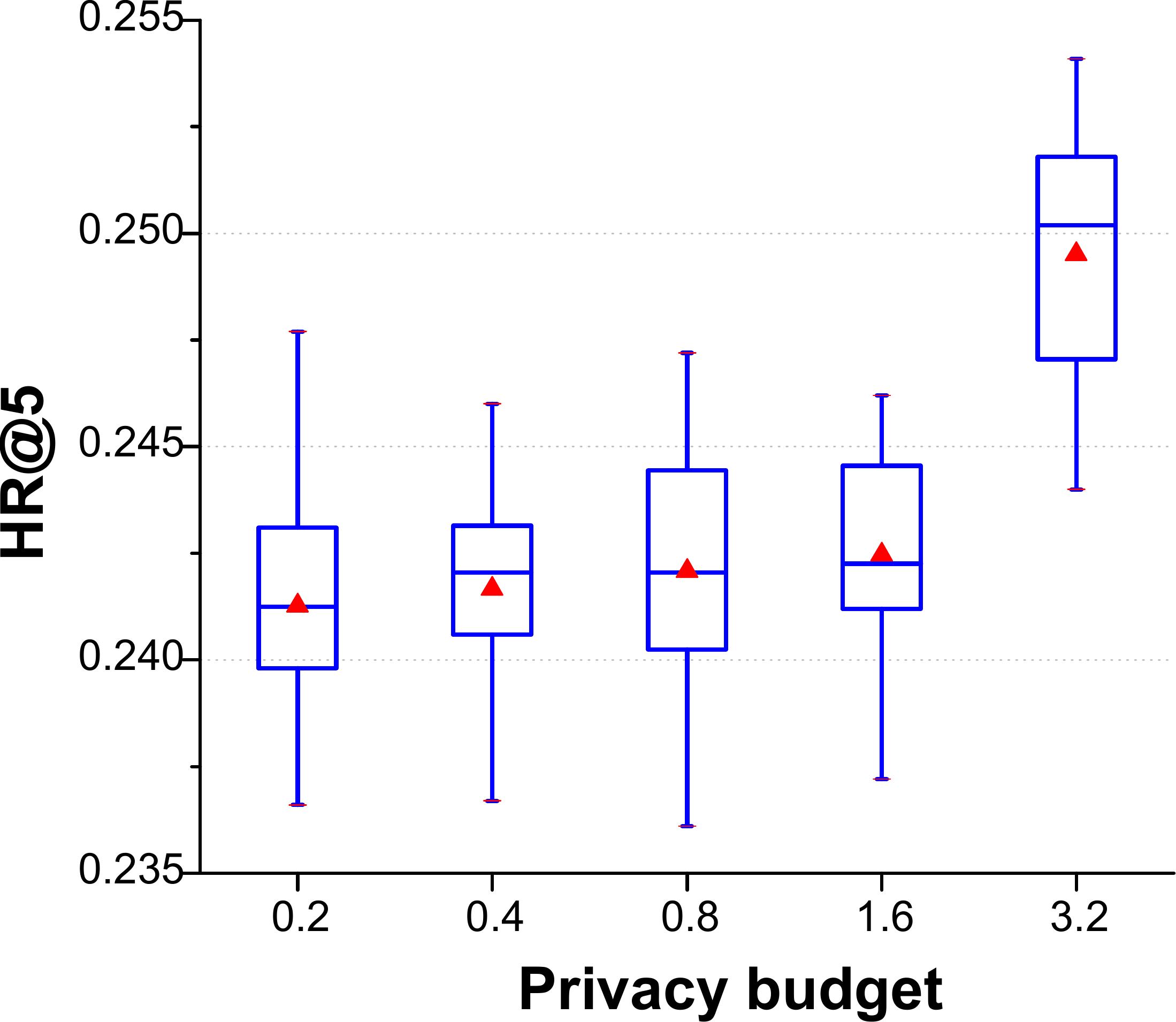}
        \label{fig7-4}
    }
    \null\hfill
    \caption{HR@k of SPIREL corresponding to varying privacy budgets}
    \label{fig7}
\end{figure*}

\begin{figure*}[!t]
    \centering
    \null\hfill
    \subfloat[Gowalla]{
        \includegraphics[scale=0.10]{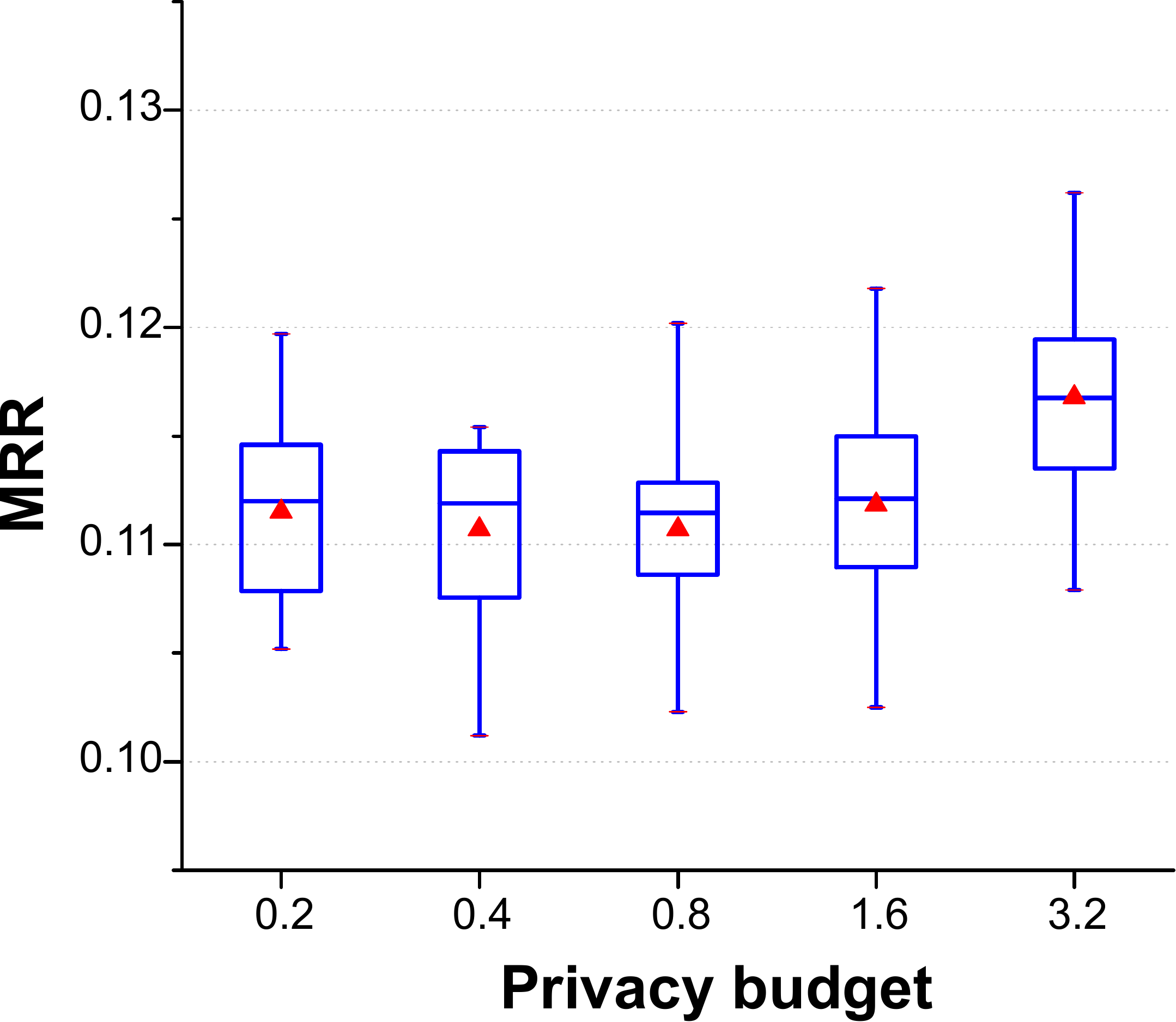}
        \label{fig8-1}
    }
    \hfill
    \subfloat[Taxi]{
        \includegraphics[scale=0.10]{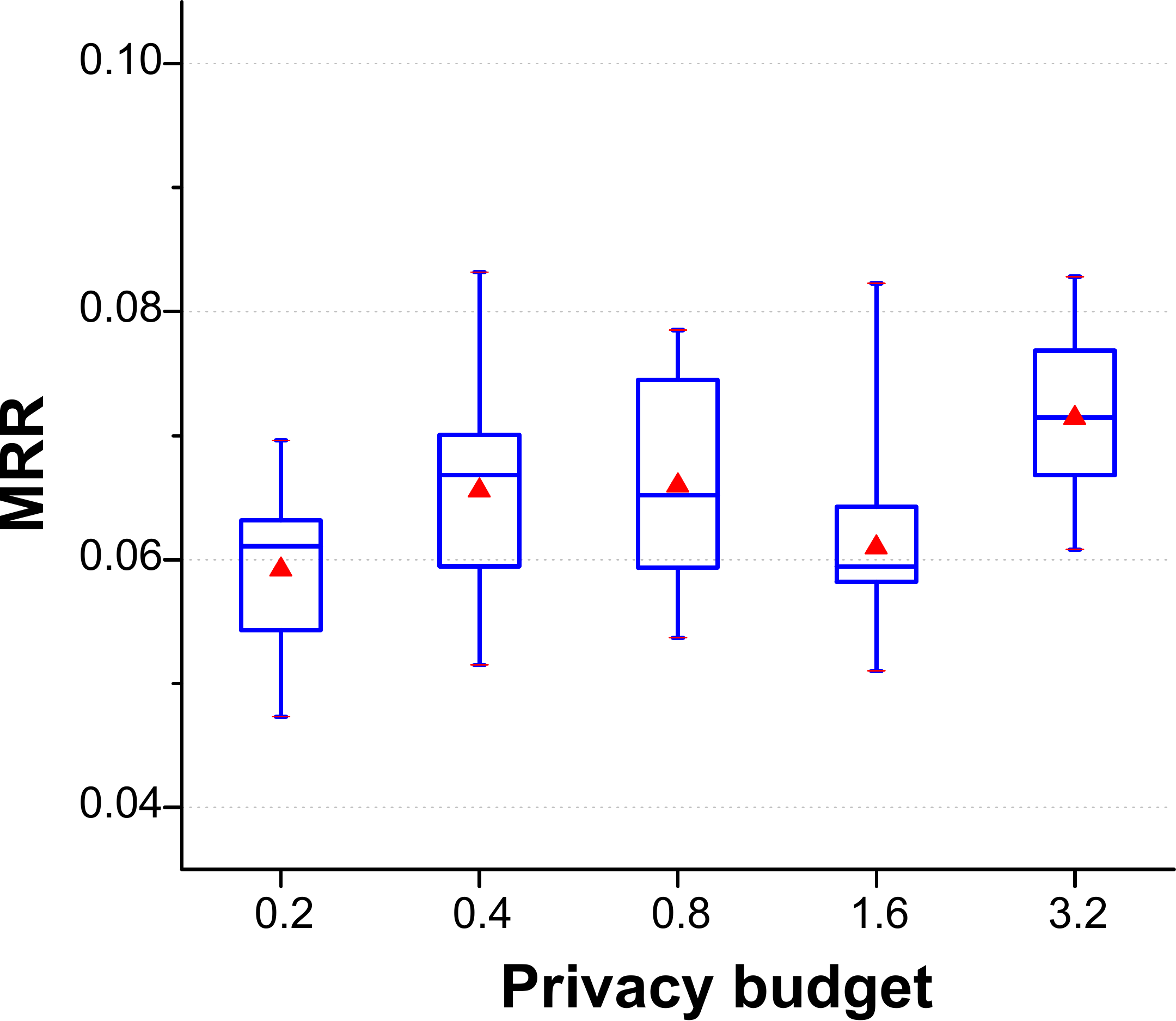}
        \label{fig8-2}
    }
    \hfill
    \subfloat[Yelp]{
        \includegraphics[scale=0.10]{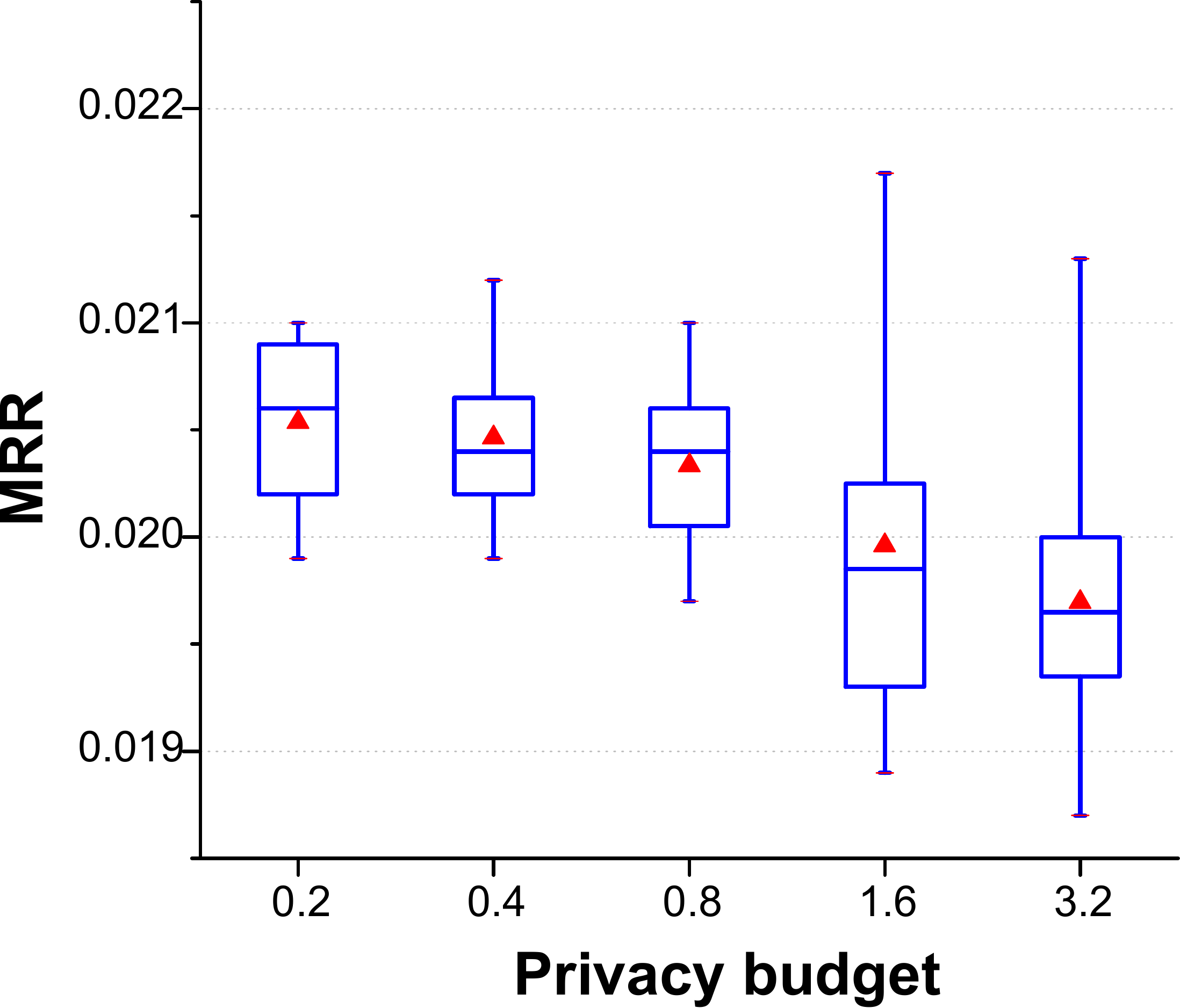}
        \label{fig8-3}
    }
    \hfill
    \subfloat[Foursquare]{
        \includegraphics[scale=0.10]{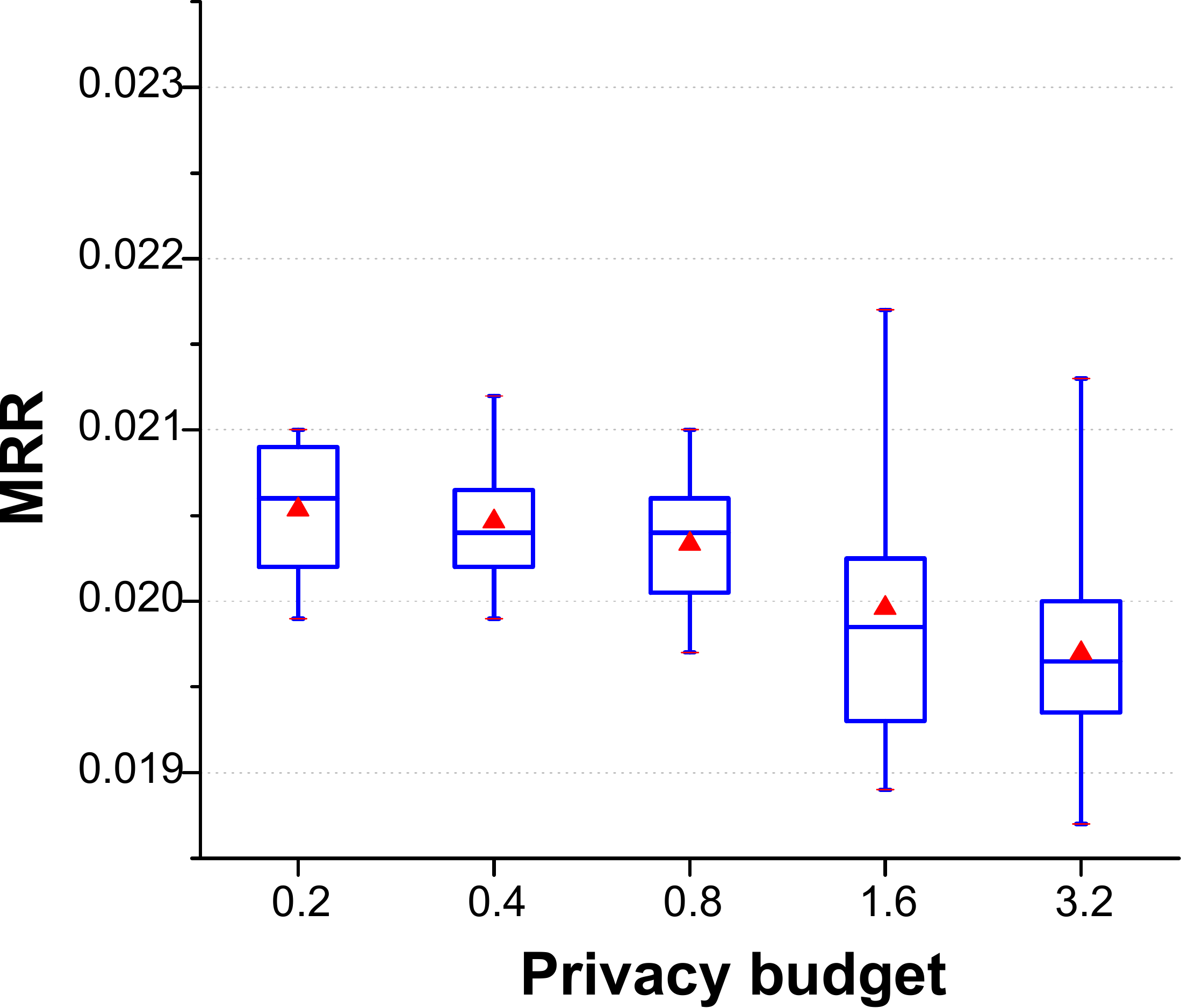}
        \label{fig8-4}
    }
    \null\hfill
    \caption{MRR of SPIREL corresponding to varying privacy budgets}
    \label{fig8}
\end{figure*}

\subsubsection{Varying privacy budgets}
\label{varying privacy budgets}

We explored the effect of privacy budget $\varepsilon$ on SPIREL. Figs. \ref{fig7} and \ref{fig8} illustrate the HR@$k$ and MRR of SPIREL over various privacy budgets ranging from 0.2 to 3.2 using boxplot visualization. Each boxplot provides six pieces of information: the maximum value (highest point), 75 percentile (upper hinge of the box), median (line inside the box), 25 percentile (lower hinge of the box), minimum value (lowest point), and arithmetic mean (triangular point). We predominantly focused on the arithmetic mean value of HR@$k$ and MRR. Overall, HR@$k$ improves as $\varepsilon$ increases because fewer noises are added to both transition patterns and gradients. On datasets \textit{Yelp} and \textit{Foursuqare}, the MRR of SPIREL does not increase as $\varepsilon$ rises. A possible explanation is that these two datasets are highly sparse and have a larger POI domain, which makes SPIREL hard to infer the relative ranking between top-k candidates from the preference scores.

\begin{figure*}[!t]
    \centering
    \null\hfill
    \subfloat[Gowalla]{
        \includegraphics[scale=0.10]{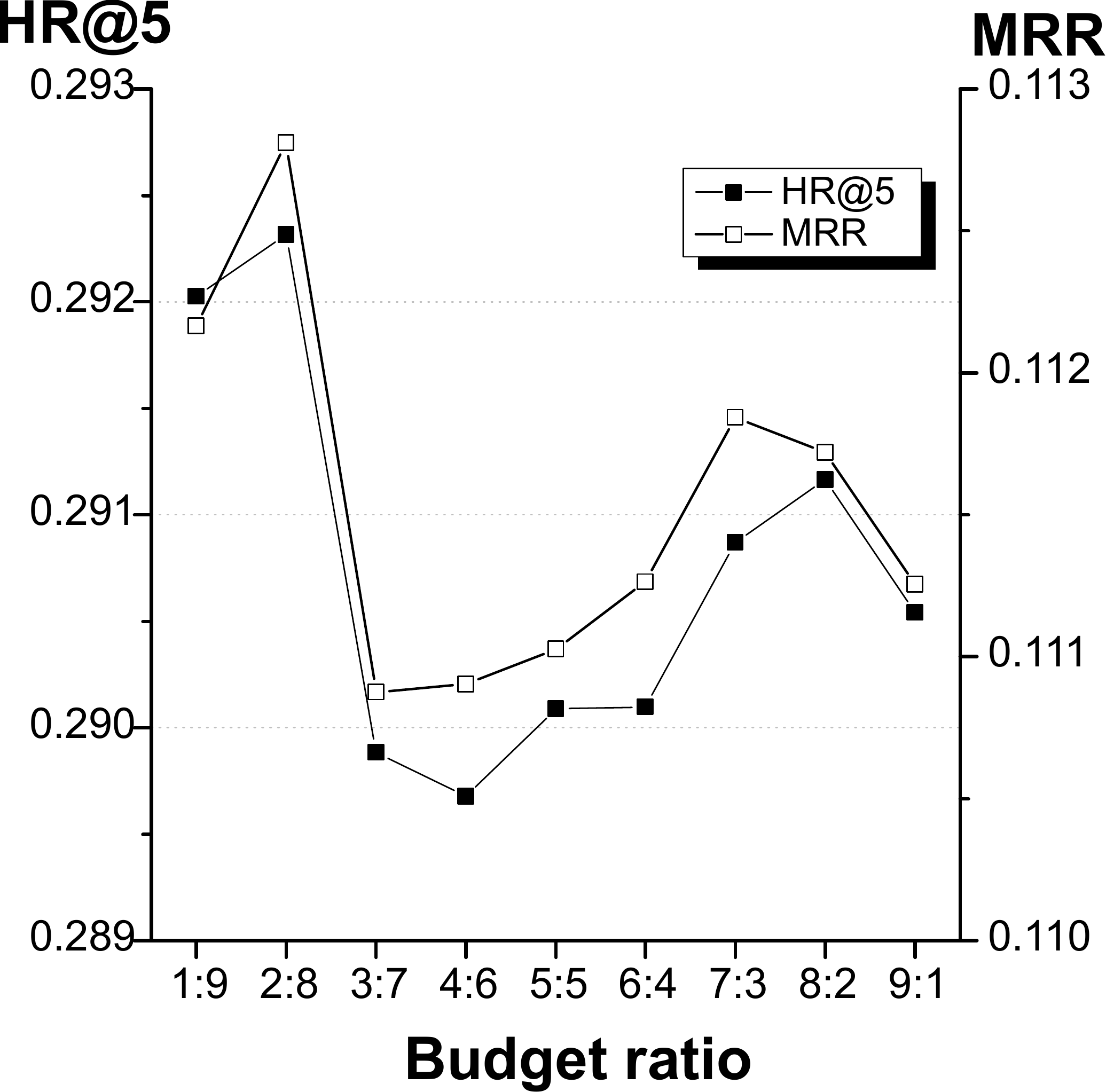}
        \label{fig9-1}
    }
    \hfill
    \subfloat[Taxi]{
        \includegraphics[scale=0.10]{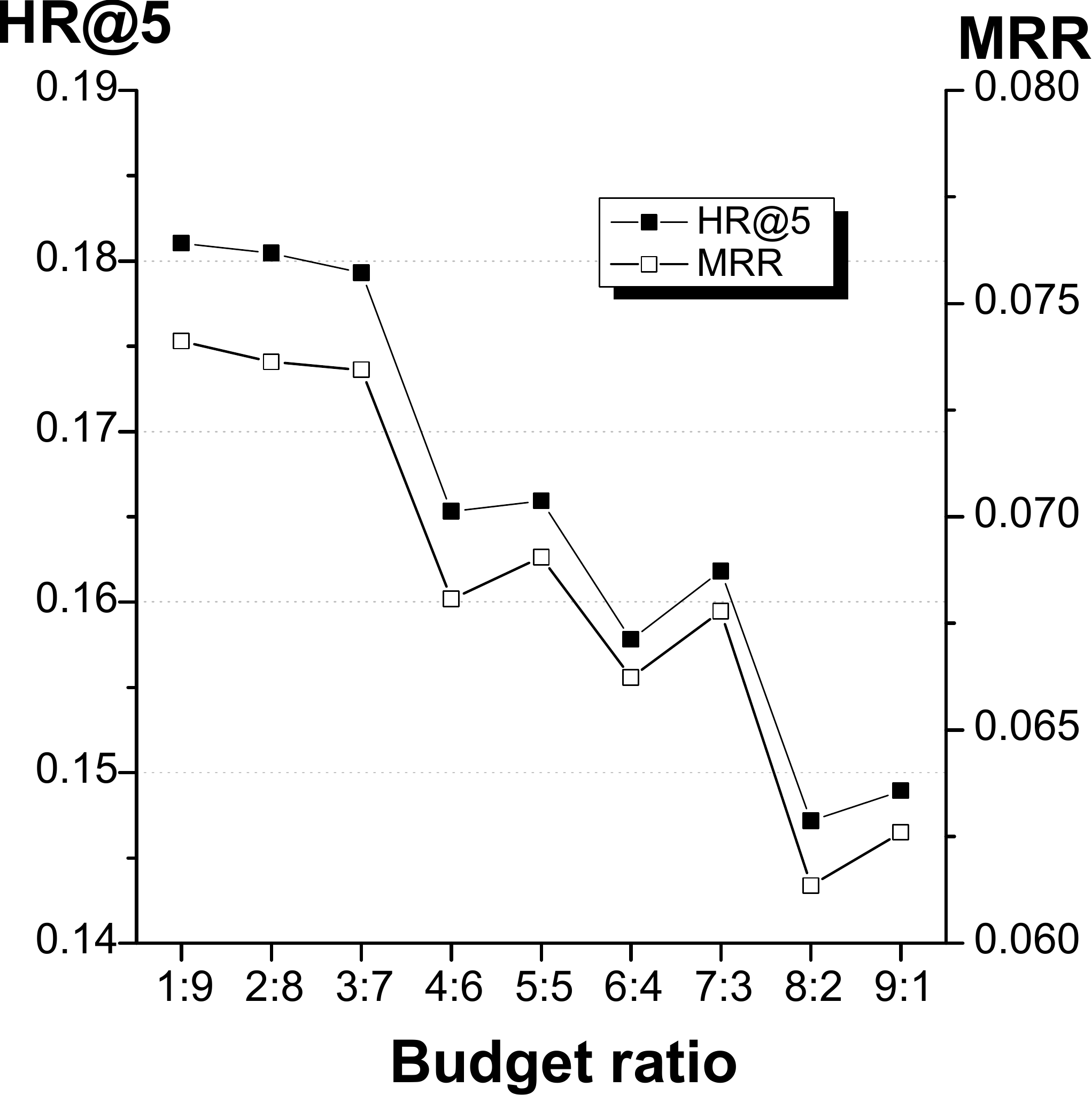}
        \label{fig9-2}
    }
    \hfill
    \subfloat[Yelp]{
        \includegraphics[scale=0.10]{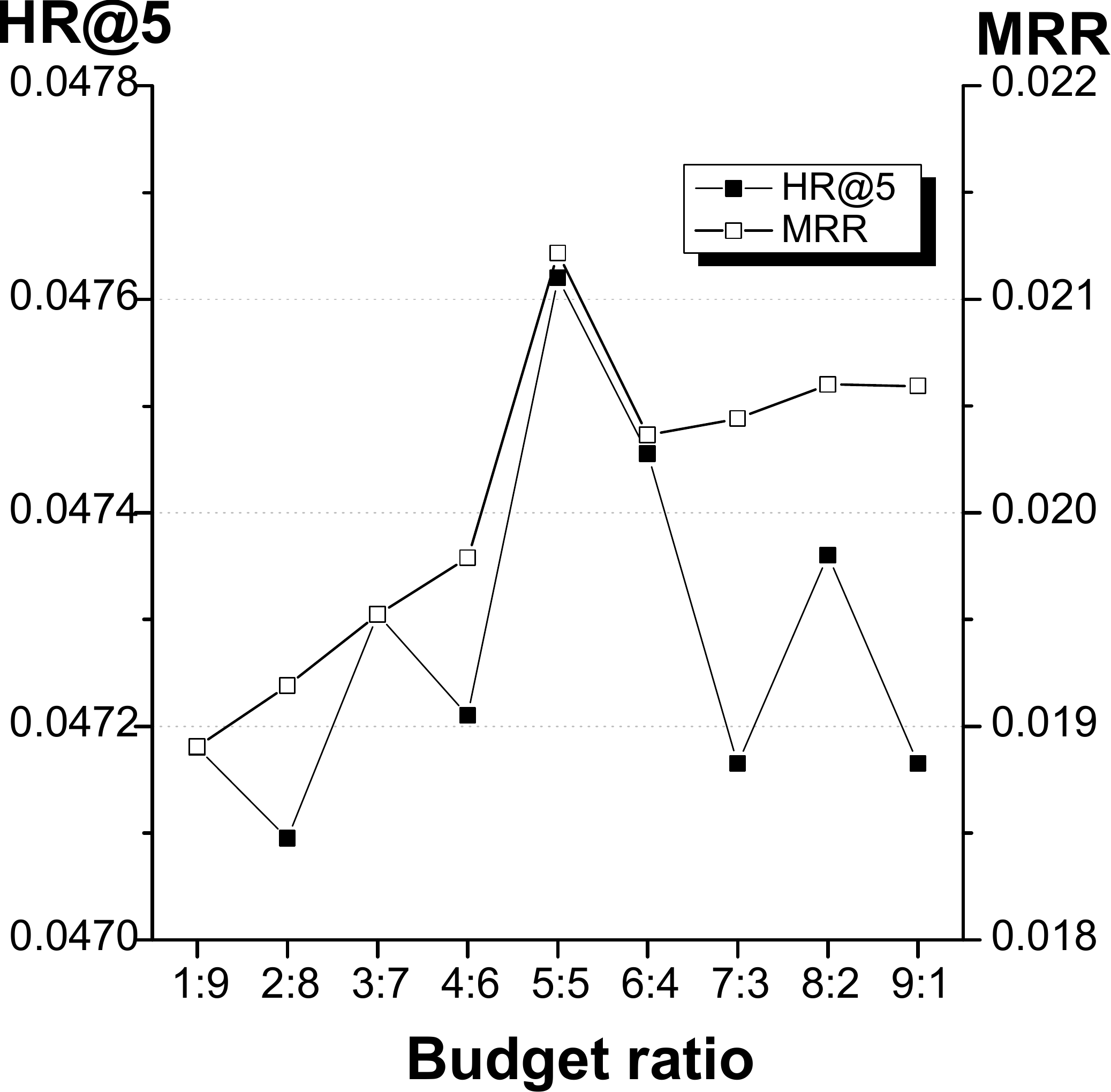}
        \label{fig9-3}
    }
    \hfill
    \subfloat[Foursquare]{
        \includegraphics[scale=0.10]{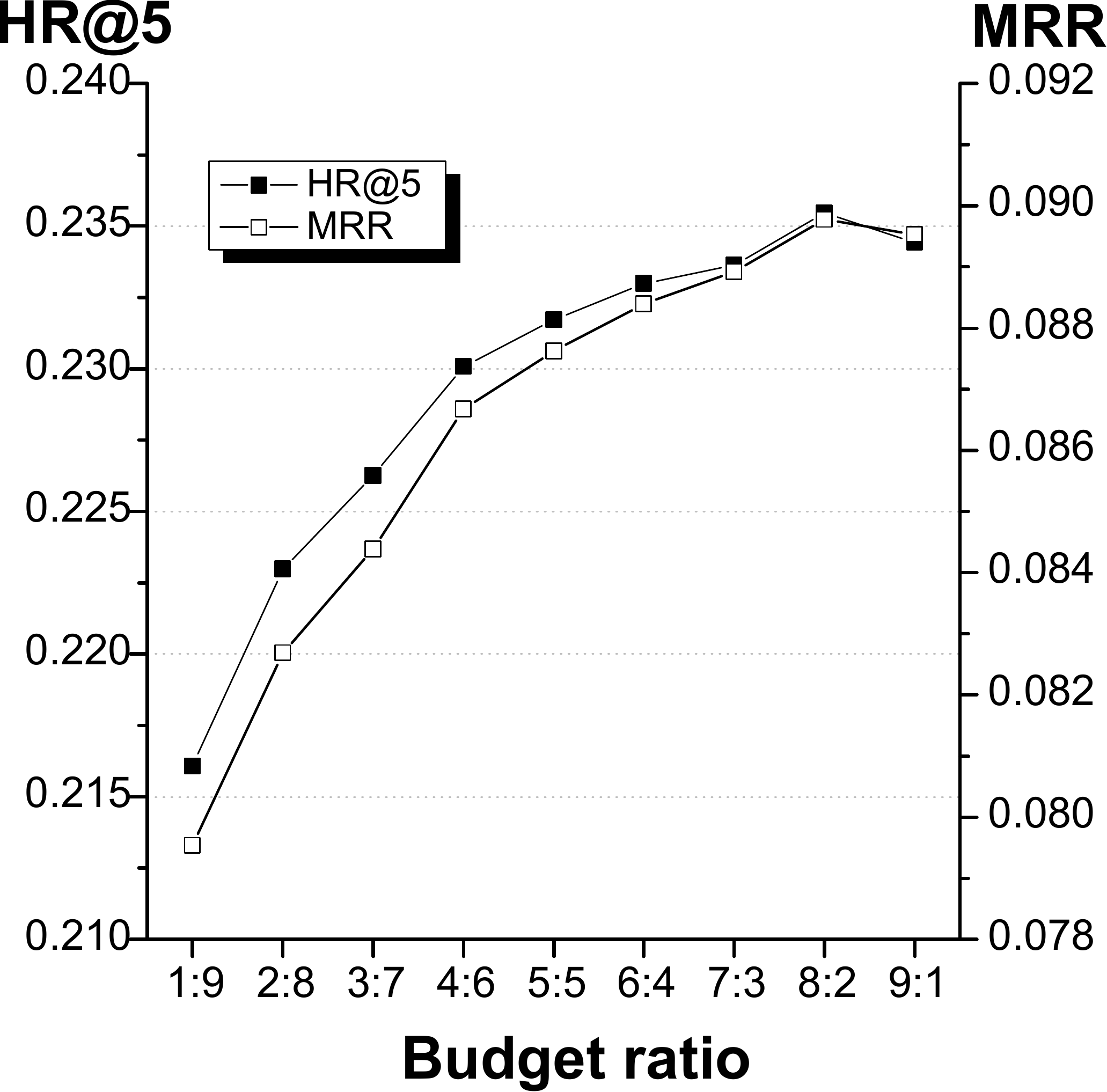}
        \label{fig9-4}
    }
    \null\hfill
    \caption{Performance of SPIREL corresponding to varying privacy budget ratios}
    \label{fig9}
\end{figure*}

\subsubsection{Varying privacy budget allocation ratios}
\label{varying privacy budget allocation ratios}

As described in Subsection \ref{transition pattern aggregation under LDP} and \ref{gradient perturbation}, each participant of SPIREL should perturb his/her transition patterns and gradients, respectively. Figure \ref{fig9} shows how the performance of the SPIREL is affected by varying the privacy budget allocation ratios from 1:9 to 9:1 (transition pattern perturbation:gradient perturbation). We observed that the HR@$k$ and MRR show a similar trend for various allocation ratios. However, these trends behave very differently for each dataset. We can conclude that when dealing with a relatively small POI domain (\textit{Gowalla} and \textit{Taxi}), focusing on the gradient perturbation results in higher recommendation performance. On the other hand, SPIREL could benefit from allocating more privacy budget to transition perturbation when the dataset is sparse (\textit{Yelp} and \textit{Foursquare}). It is the reason why the performance gain of SPIREL-PM stands out in Figure \ref{fig5-4}, where the other datasets have less room for performance to rise thanks to the piecewise mechanism when equally allocating the privacy budget.

\begin{figure*}[!t]
    \centering
    \null\hfill
    \subfloat[Gowalla]{
        \includegraphics[scale=0.10]{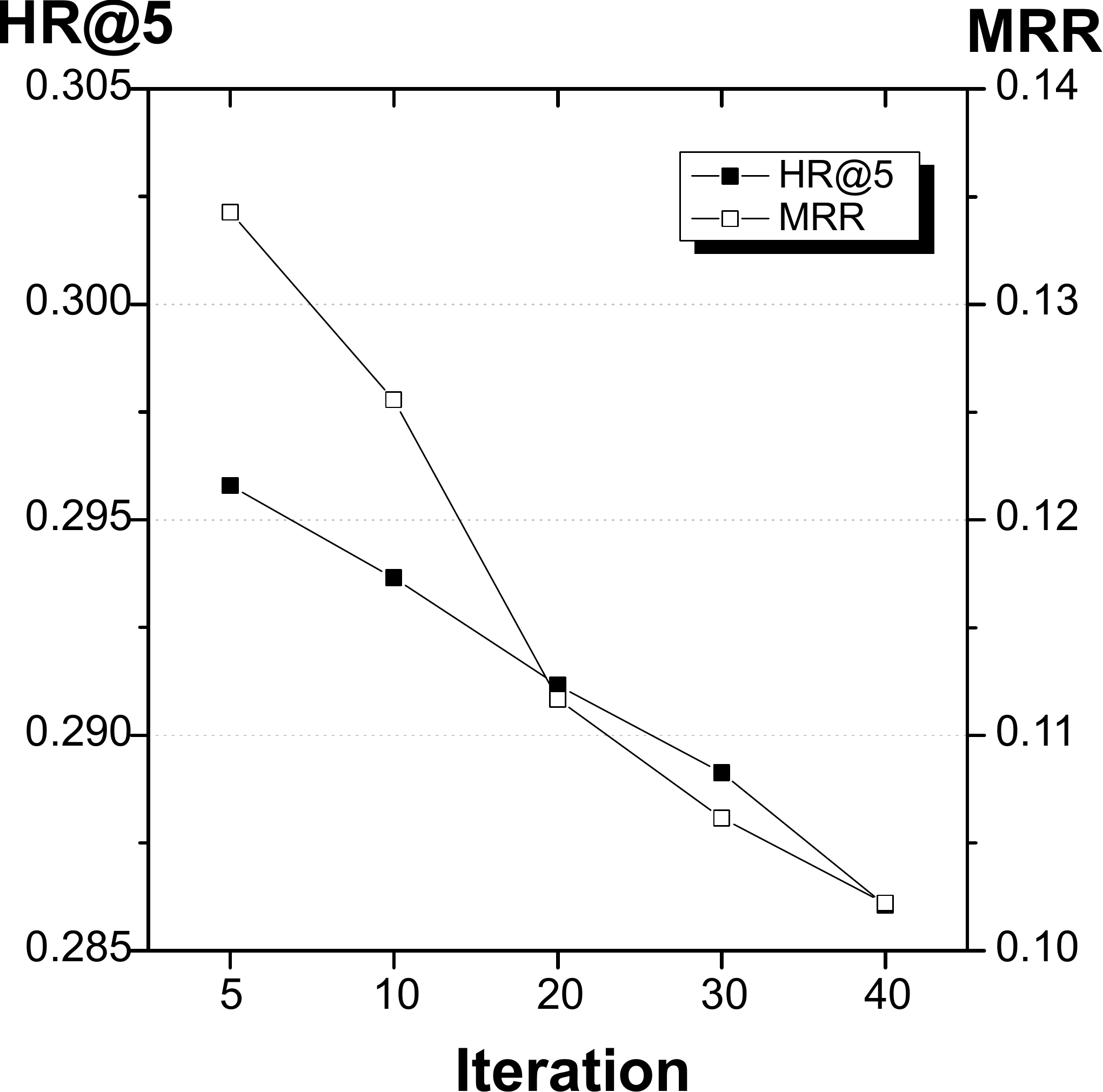}
        \label{fig10-1}
    }
    \hfill
    \subfloat[Taxi]{
        \includegraphics[scale=0.10]{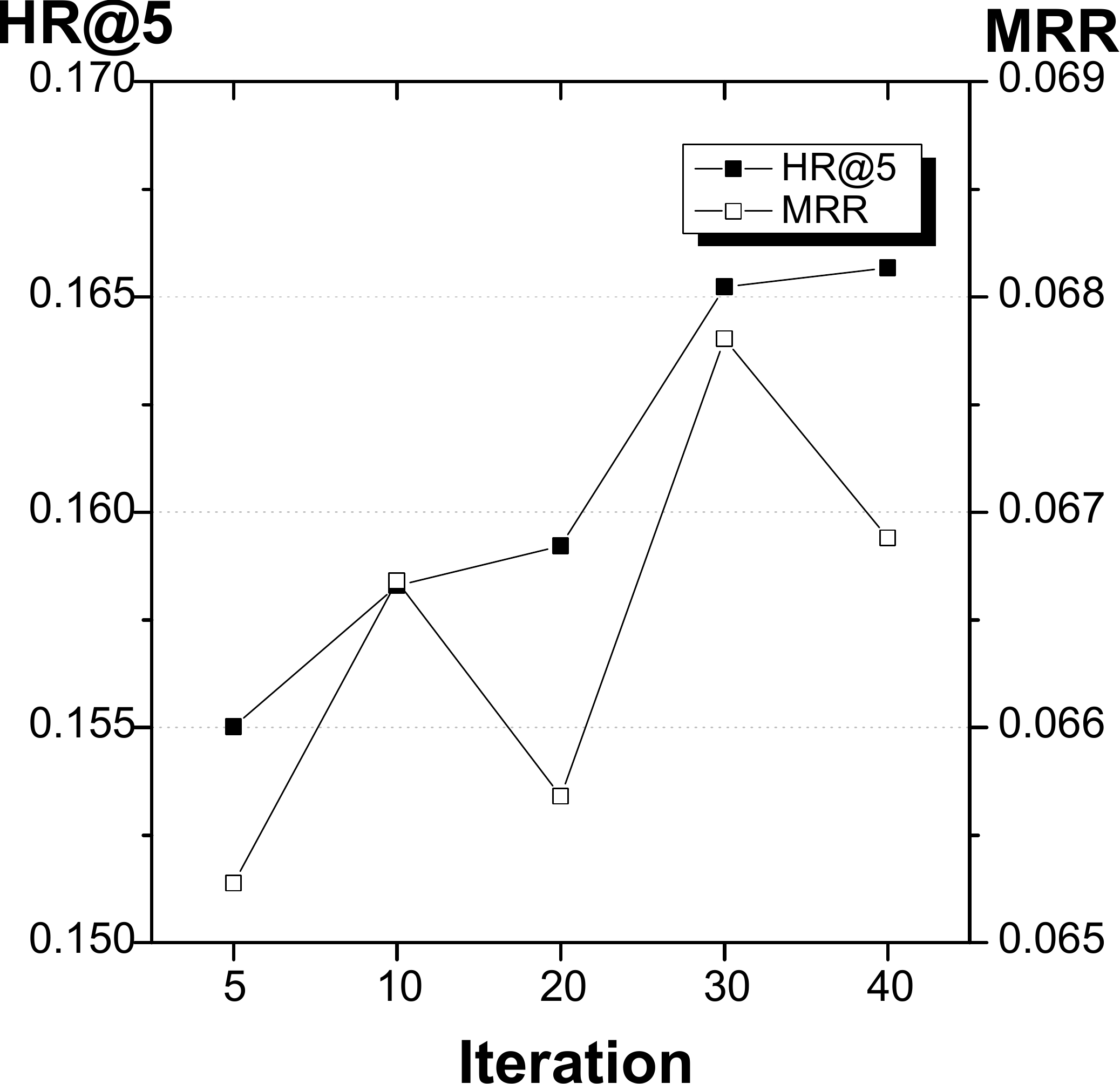}
        \label{fig10-2}
    }
    \hfill
    \subfloat[Yelp]{
        \includegraphics[scale=0.10]{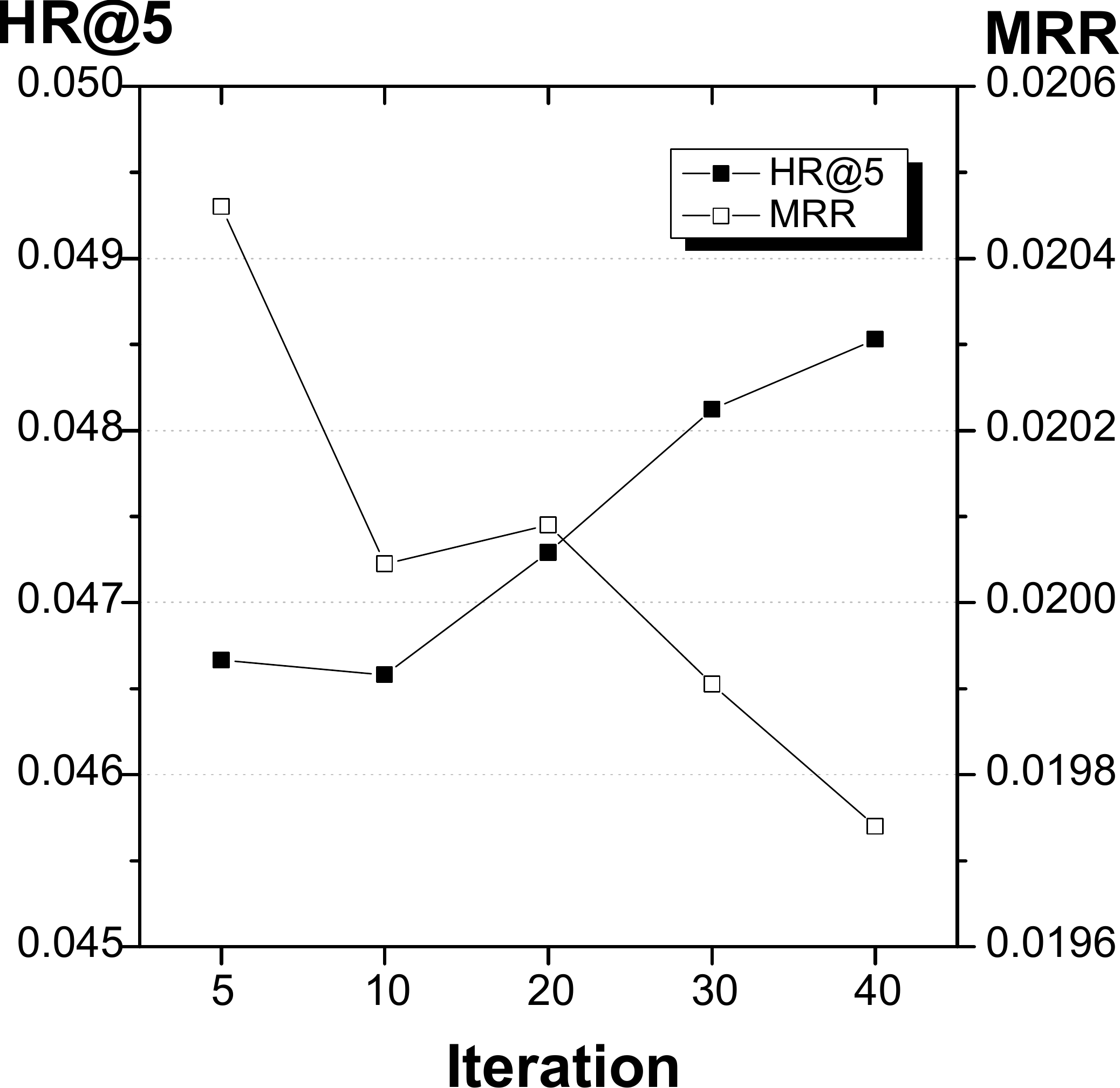}
        \label{fig10-3}
    }
    \hfill
    \subfloat[Foursquare]{
        \includegraphics[scale=0.10]{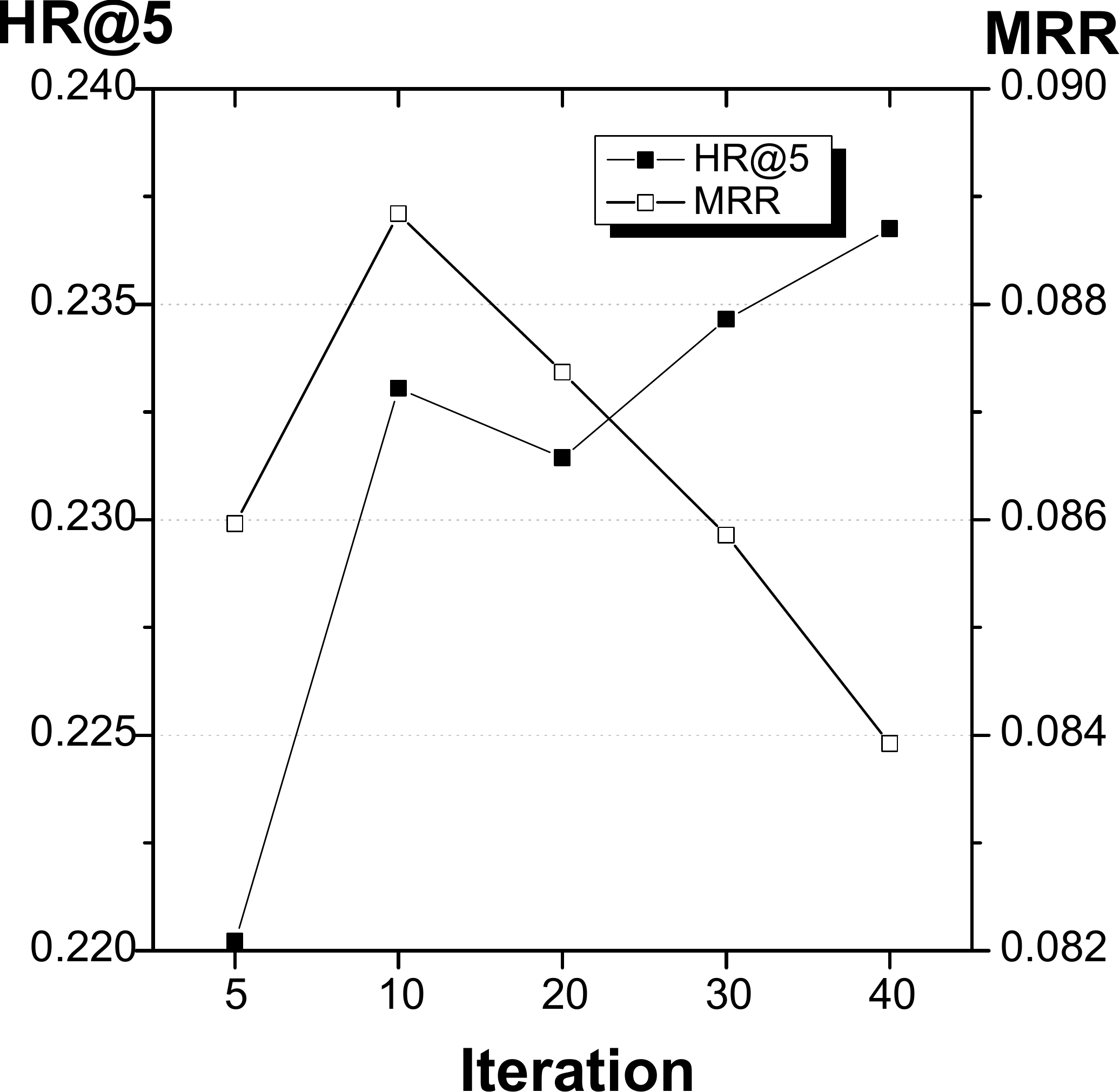}
        \label{fig10-4}
    }
    \null\hfill
    \caption{Performance of SPIREL corresponding to varying numbers of maximum iterations}
    \label{fig10}
\end{figure*}

\begin{figure*}[!t]
    \centering
    \null\hfill
    \subfloat[Gowalla]{
        \includegraphics[scale=0.10]{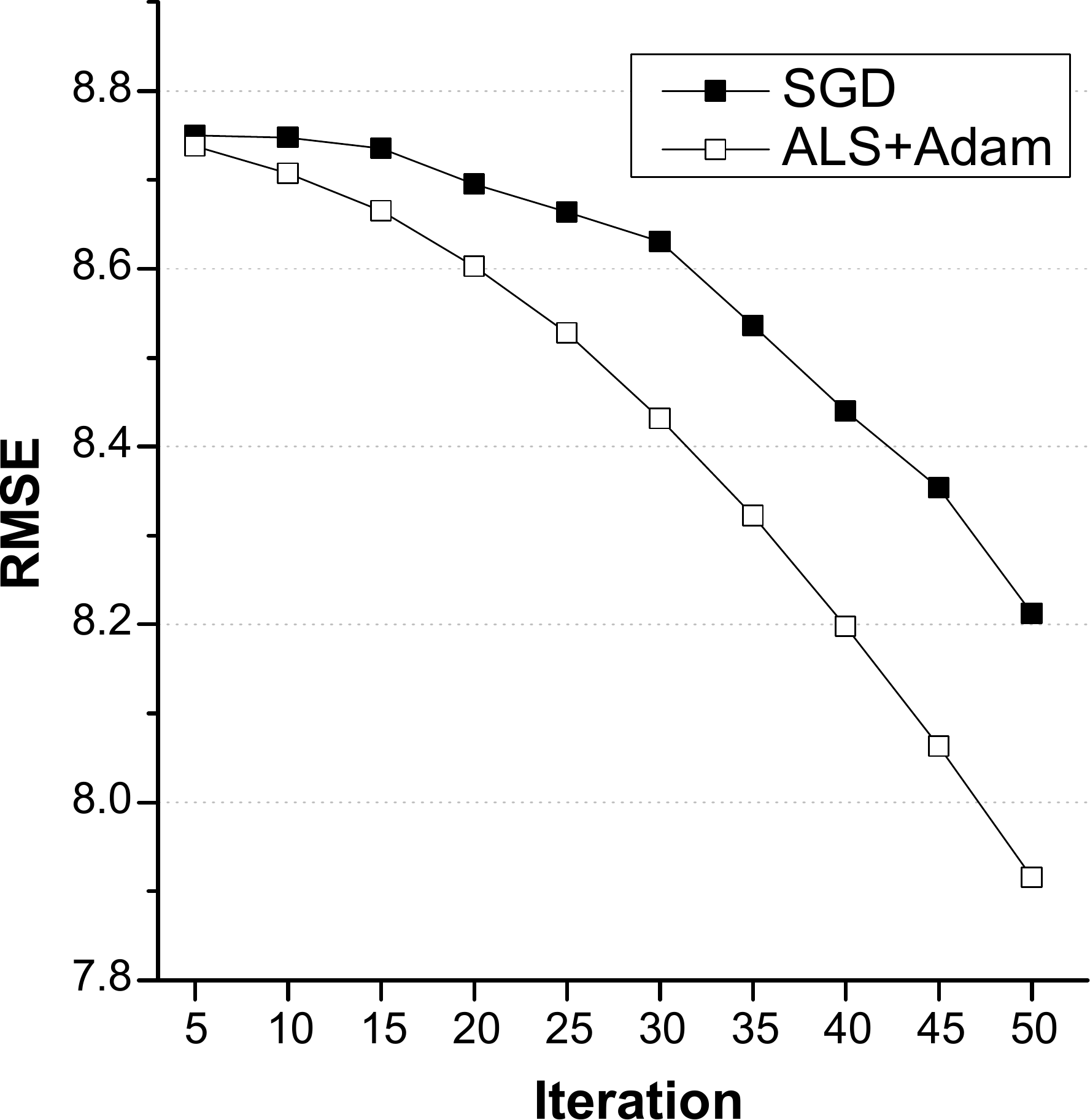}
        \label{fig11-1}
    }
    \hfill
    \subfloat[Taxi]{
        \includegraphics[scale=0.10]{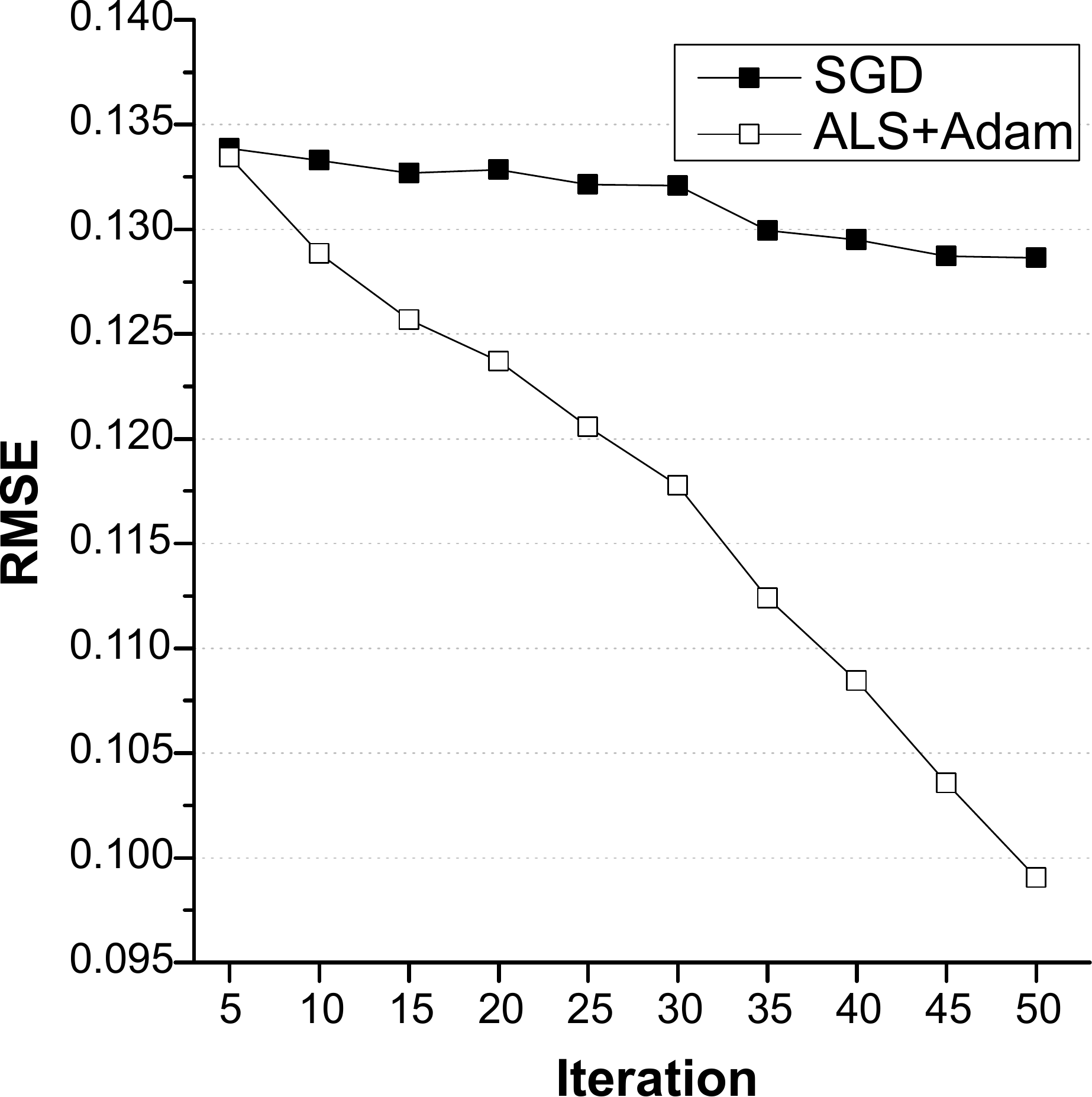}
        \label{fig11-2}
    }
    \hfill
    \subfloat[Yelp]{
        \includegraphics[scale=0.10]{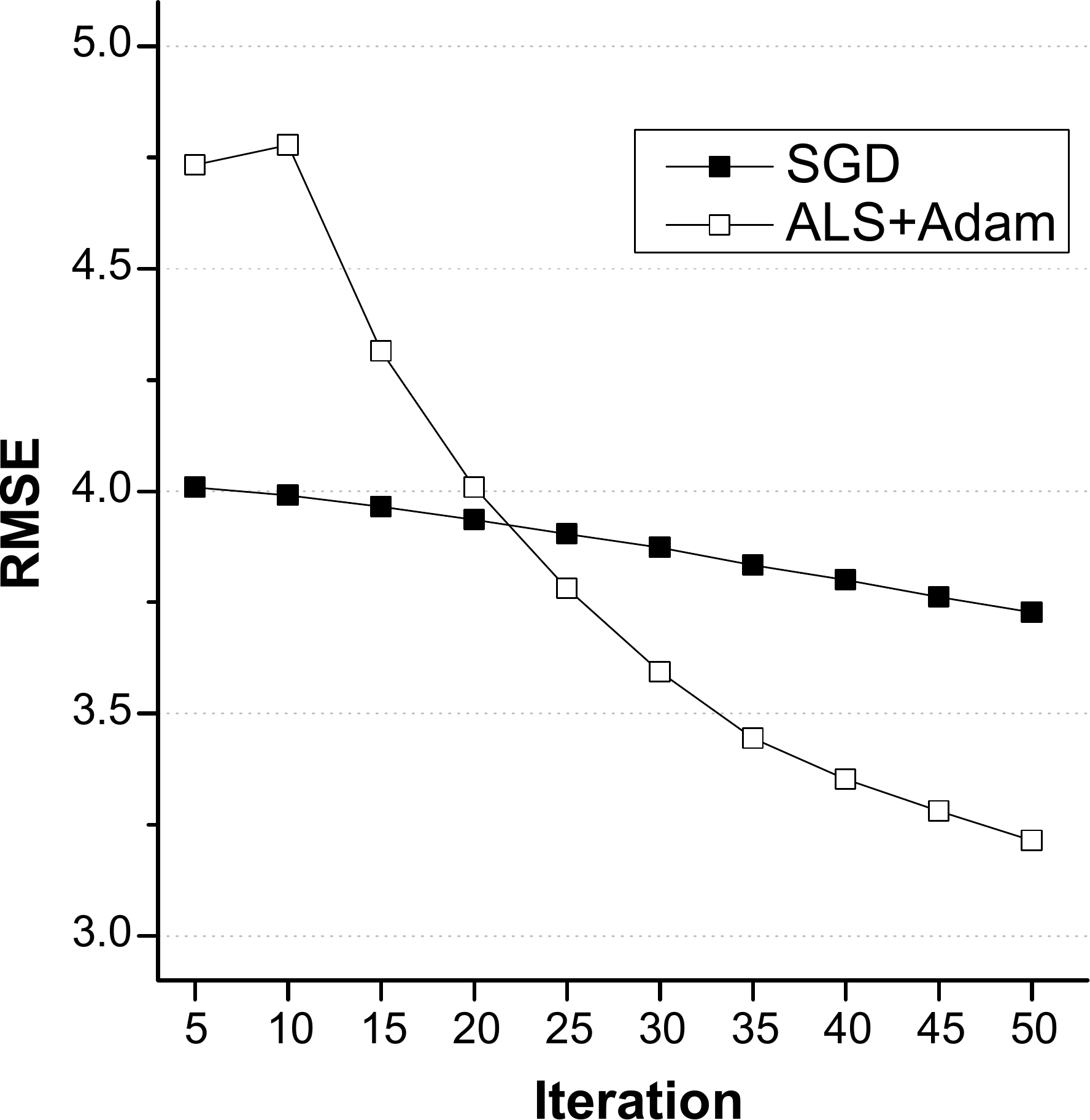}
        \label{fig11-3}
    }
    \hfill
    \subfloat[Foursquare]{
        \includegraphics[scale=0.10]{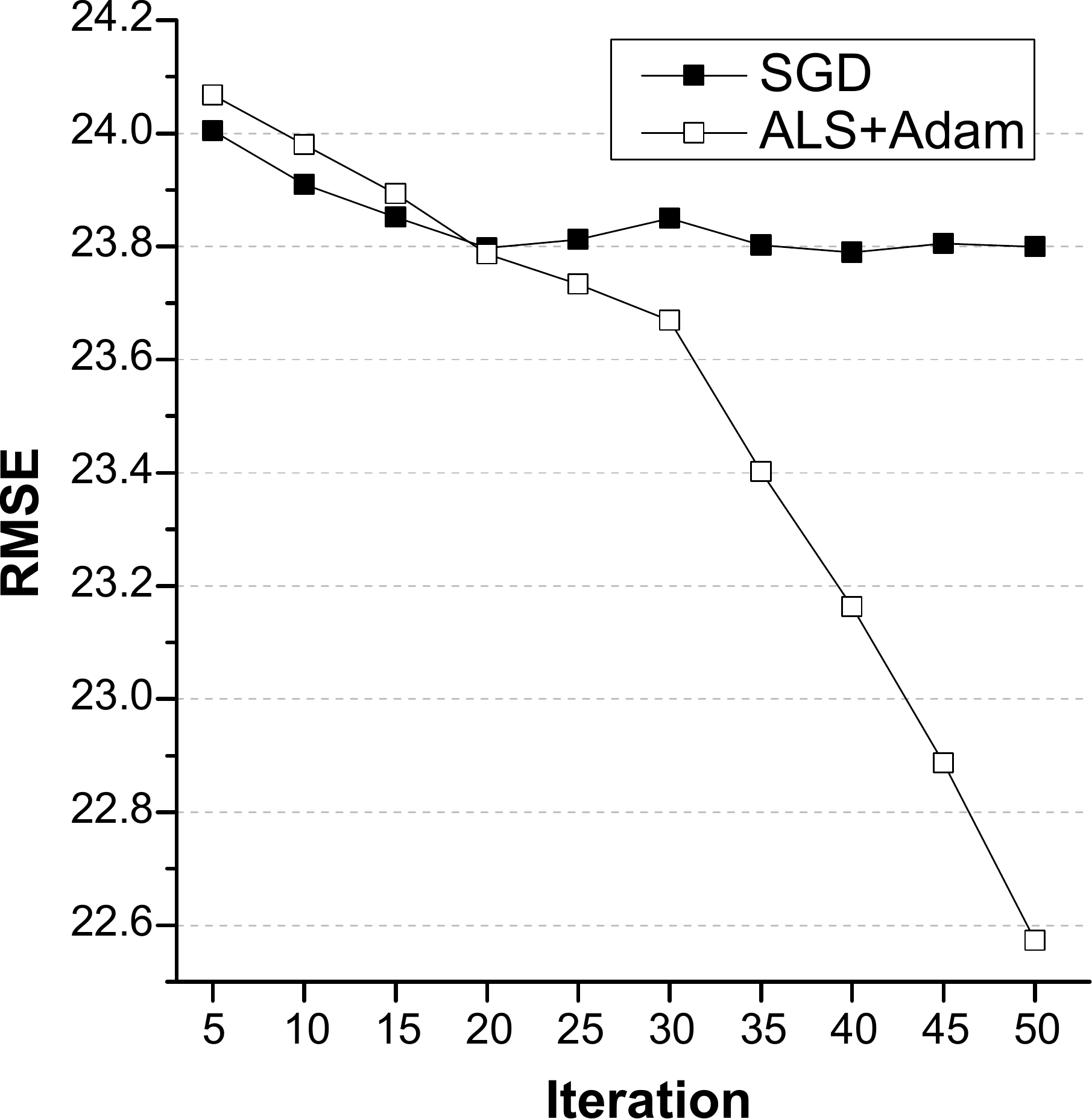}
        \label{fig11-4}
    }
    \null\hfill
    \caption{Root mean square errors of SPIREL over 50 iterations with or without ALS and Adam optimizer}
    \label{fig11}
\end{figure*}

\subsubsection{Varying numbers of maximum iterations}
\label{Varying numbers of maximu iterations}

The maximum number of iterations is limited in privacy-preserving recommendation systems, as each iteration consumes a portion of the privacy budget. Thus, we should consider both the stability and convergence of learning process, otherwise SPIREL would suffer from underfitting or even cause the loss function to diverge. We analyzed the effects of the maximum number of iterations on SPIREL in Figure \ref{fig10}. Moreover, we illustrate the root mean square error (RMSE) of loss function during 50 iterations to demonstrate the effects of using both ALS and Adam optimizer in Figure \ref{fig11}. 

While the RMSE consistently decreases in the dataset \textit{Gowalla}, the performance of SPIREL decreases as the maximum number of iterations increases. We can conclude that SPIREL suffers from the overfitting problem, and this indicates that SPIREL can reach optimal within five iterations for small dataset. Except for the dataset \textit{Gowalla}, the HR@$k$ of SPIREL consistently increases during the learning process. Again, due to the large POI domain size of the datasets \textit{Yelp} and \textit{Foursquare}, SPIREL experiences performance drops in MRR despite HR@$k$ increases.

\begin{table*}[t]
\centering
\caption{HR@5 and MRR of SPIREL with SGD and SPIREL with ALS+Adam}
\label{table3}
\begin{tabular}{|c|c|c|c|c|c|c|c|c|}
\hline \centering 
Dataset & \multicolumn{2}{|c|}{Gowalla} & \multicolumn{2}{c|}{Taxi} & \multicolumn{2}{c|}{Yelp} & \multicolumn{2}{c|}{Foursquare}\\
\hline 
\hline \centering 
Metric & HR & MRR & HR & MRR & HR & MRR & HR & MRR \\
\hline \centering 
SGD & 0.0101 & 0.0040 & 0.0122  & 0.0049 & 0.0056 & 0.0022 & 0.0031 & 0.0012 \\
\hline \centering 
ALS+Adam & 0.2983 & 0.1311 & 0.1569 & 0.0662 & 0.0473 & 0.0201 & 0.2321 & 0.0874 \\
\hline
\end{tabular}
\end{table*}

Finally, we examined the effects of combining ALS and Adam optimizer compared to naive SGD updates. The results are shown in Table \ref{table3}. SPIREL obtains a huge performance improvement in all datasets when using our learning framework. We attribute the good performance of SPIREL to merging the advantages of ALS and Adam. We also observed that increasing the learning rate of SGD does not improve the performance because the loss function diverges immediately.

\Figure[ht]()[width=3.1in]{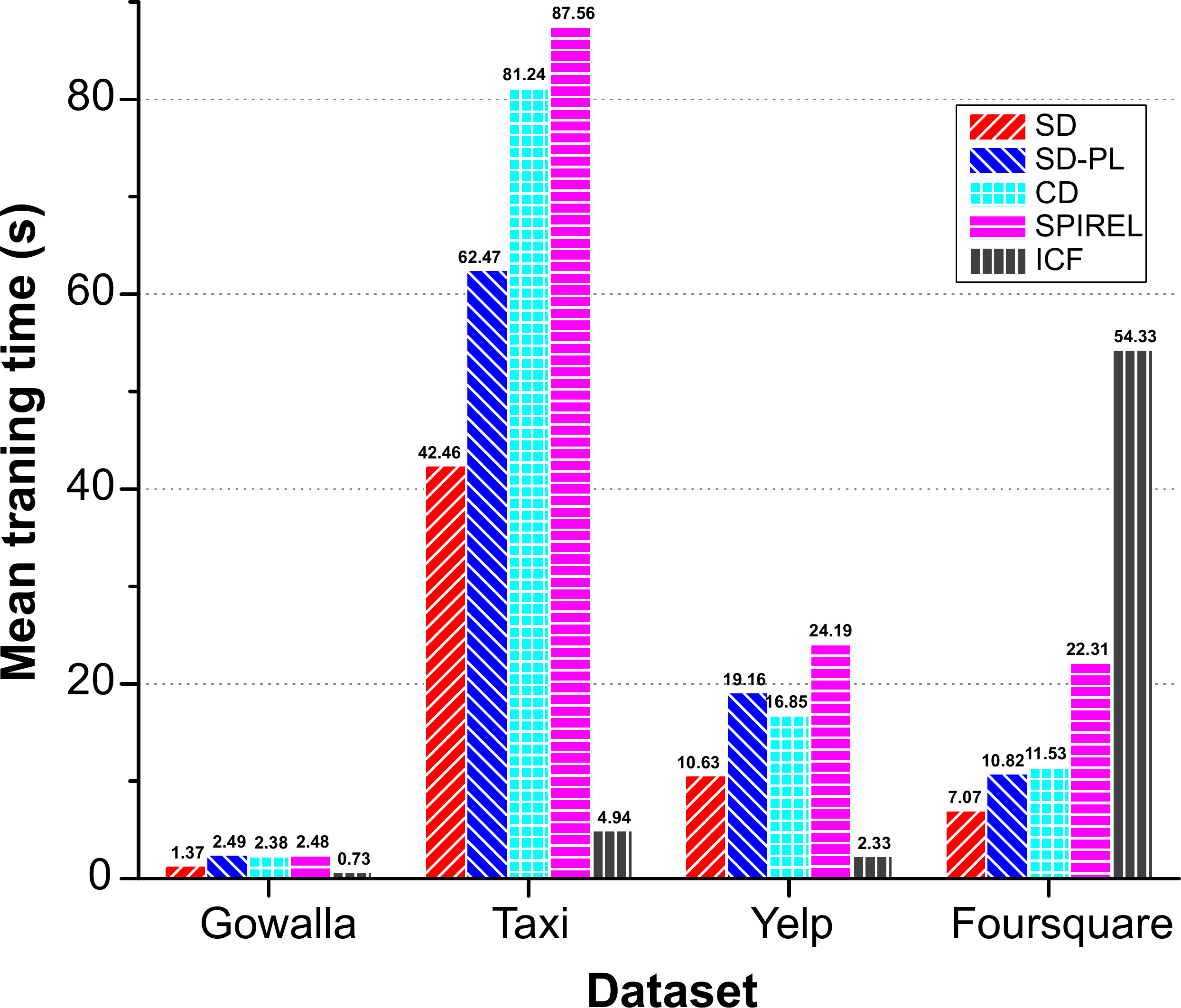}
{Training time comparison on the different datasets \label{fig12}}

\subsubsection{Training time comparison}
\label{model trainign time comparison}

Because of the vast numbers of users and items in collaborative filtering, the recommender model would be inefficient if it takes too much time for training. So, we compared the training time of five algorithms. The training time is reported in Figure \ref{fig12}. Note that the training time includes server-side computations as well as user-side computations. Thus, the actual training time in a real-world setting is less than the results shown in Figure \ref{fig12}. We fixed the maximum number of iterations to 20 for MF-based algorithms. 

We observed that SPIREL required the longest training time across all datasets except for the \textit{Foursquare} dataset. This can be explained by the computation time for building the POI-POI matrix for transfer learning. Overall, the privacy-preserving approaches (SD-PL, SPIREL) demonstrated a longer training time than their non-private versions (SD, CD) because of the time for perturbation. Although ICF showed the shortest training time on \textit{Gowalla}, \textit{Taxi}, and \textit{Yelp} datasets, its training time is sensitive to the domain size of POIs due to the nature of comparing every POI-POI pairs to derive the similarity scores. Accordingly, it required the longest training time on the \textit{Foursquare} dataset. 


\section{Related work}
\label{related work}

\begin{table}
\centering
\caption{Comparison between SPIREL and other privacy-preserving recommendation methods}
\label{table4}

\begin{tabular}{|c|c|c|c|c|c|}
\hline
Group & Method & DP model & Learning model & Data type\\
\hline
\hline
CD & SPIREL & LDP & MF & implicit \\
\hline
\multirow {3}{*}{SD} & \cite{hua2015differentially}, \cite{zhang2018probabilistic} & CDP & MF & explicit\\
& \cite{shin2018privacy} & LDP & MF & explicit\\
& \cite{guo2019locally} & LDP & Item-based CF & implicit\\
\hline
\end{tabular}

\end{table}

The problem of successive POI recommendation has received significant attention recently \cite{cheng2013you,liu2013personalized,lian2014geomf,chang2018content}. To predict where a user will visit next, we have to consider the relationship between POIs in addition to that between users and POIs. However, the traditional MF-based recommendation methods neglect the sequential patterns and infer the user preferences only based on the places the users visited. To overcome this problem, \cite{rendle2010factorizing, feng2015personalized} consider the sequential influences by integrating the Markov chain into MF and obtain good performance on next-item recommendation. Currently, geographical influence is fused with the MF algorithm to enhance the accuracy of POI recommendations \cite{lian2014geomf, rahmani2019lglmf}.

Our research direction was to incorporate the relationship between POIs by adapting the transfer learning approach \cite{pan2010transfer, pan2011transfer, pan2013transfer}. Most transfer learning methods under MF technique utilize relevant data from other domains into the target domain by sharing the latent matrix. For example, the approach called \textit{coordinate system transfer} \cite{pan2010transfer} first learns the user latent matrix $U_{A}$ from an auxiliary domain and generates the user latent matrix $U_{T}$ for target domain based on $U_{A}$. Further, an extension of this approach was proposed that exploits the implicit binary feedback (e.g., like/dislike) on items to construct a target explicit rating matrix (e.g., 5-star rating) \cite{pan2013transfer}.

DP \cite{dwork2006calibrating} is a rigorous privacy standard; the output of a DP mechanism should not reveal much information specific to any individual. DP in a centralized setting requires a trusted data curator who collects original data from users. Recently, a local version of DP (LDP) was proposed. In the local setting, each user perturbs his/her data and sends the perturbed data to the data curator. Since the original data never leaves the device of the users, the LDP mechanisms have the benefit of not requiring any trusted data curator. Accordingly, many companies attempt to adopt LDP when collecting data from clients for privacy reasons \cite{erlingsson2014rappor,apple2014learning,ding2017collecting,nguyen2016collecting}.

Beyond estimating the frequency of a single item, there are several works on discovering item sequences through LDP. \cite{wang2018privtrie} found frequent new words by building a noisy prefix tree. \cite{kim2018learning} and \cite{fanti2016building} used the estimated frequency of $n$-gram to identify the frequent new terms. \cite{chen2012differentially} also used the $n$-gram model to release noisy trajectory. Our work employed the $n$-gram approach to model the POI sequences. Instead of using the estimated frequency directly, we turn the frequency into a confidence score and map it to the users' preferences.

Table \ref{table4} summarizes the important differences between the SPIREL and previous works. To the best of our knowledge, no previous MF-based recommendation systems \cite{hua2015differentially, zhang2018probabilistic, shin2018privacy} exploit the implicit feedback to recommend the next POI. Specifically, Hua et al. \cite{hua2015differentially} proposed an objective function perturbation method. In their work, a trusted data curator added Laplace noises to the objective function so that the factorized item matrix satisfies CDP. Zhang et al. \cite{zhang2018probabilistic} proposed a probabilistic MF-based recommendation system with personalized CDP. They used a random sampling method to satisfy the privacy requirements of different users. Shin et al. \cite{shin2018privacy} proposed a MF-based recommendation system under LDP. In their method, users updated their profile vectors using SGD locally and submitted the perturbed gradients to the server. They reduced the amount of noise added on gradients by adopting the dimensionality reduction technique into the item domain. In short, these works focused on decreasing the RMSE during the learning process and attempted to predict users' preferences using explicit feedback that has a fixed scale rating.

Some works adopted LDP for building recommendation systems \cite{guo2019locally, asada2019and, chen2020practical}. Guo et al. \cite{guo2019locally} proposed a privacy-preserving item-based CF technique under LDP. Their method predicted the next item based on the similarity score that is estimated through item co-occurrence frequencies. However, employing only the similarity score between items had a limit to predict the preferences of users on POIs. Asada et al. \cite{asada2019and} proposed a location preference recommendation system with LDP. However, the location preference here indicated whether a user wants to publish information about a specific location or not, which is a different area from recommending the next POI. Gao et al. \cite{chen2020practical} proposed a POI recommendation model using Factorization Machines. They estimated the user preferences by combining a linear model with a high-order feature model and adopted decentralized SGD to train the model in a privacy-preserving manner.


\section{Conclusion}
\label{conclusion}

In this paper, we proposed a novel successive POI recommendation framework under LDP, named SPIREL. We first collected noisy transition patterns between two consecutive POIs. Then, we calculated a confidence score for each transition pattern for the POI-POI matrix. Subsequently, we assisted in predicting the preferences of the next POI in the user-POI domain using the confidence score from the POI-POI domain by transfer learning approach. For preventing privacy leakage from the stage of iterative learning, we integrated LDP mechanisms to collect gradients and combined both ALS and SGD to train SPIREL. By extensive experiments on four public datasets, we demonstrated that SPIREL provides significantly better POI recommendation performance than the existing privacy-preserving recommendation methods.


\EOD

\begin{IEEEbiography}[{\includegraphics[width=1in,height=1.25in,clip,keepaspectratio]{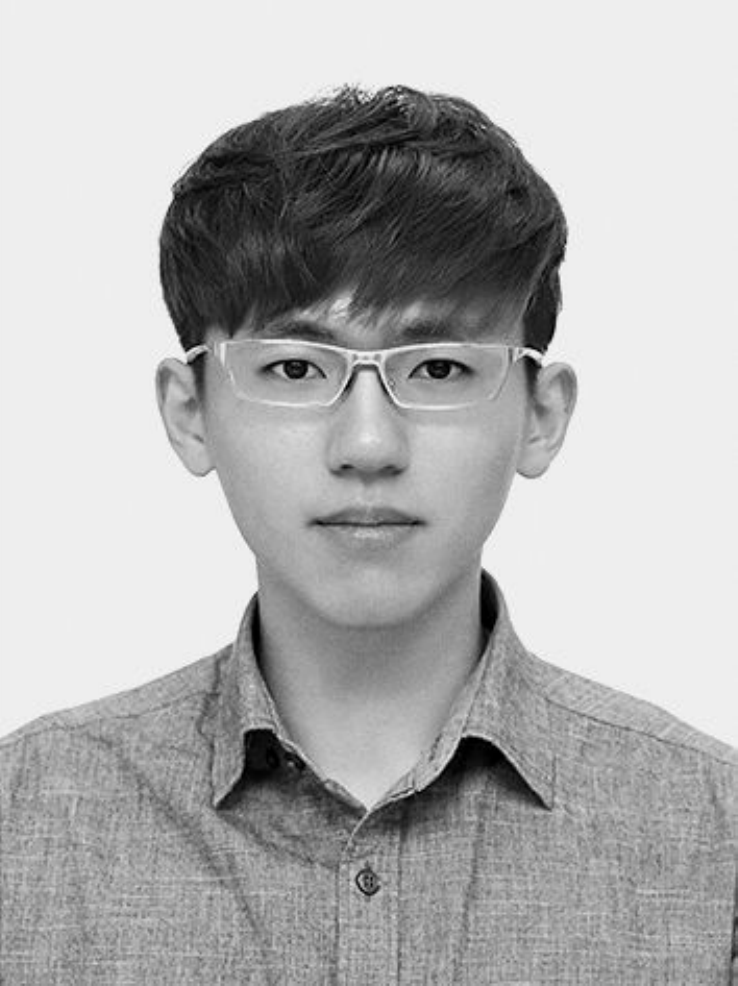}}]{\textbf{Jong Seon Kim}} received his B.S. degree in computer science from Korea University in 2016. He is currently undergoing an integrated Ph.D. program in the Department of Computer Science and Engineering in Korea University. His research interests include location privacy and recommendation systems.
\end{IEEEbiography}

\begin{IEEEbiography}[{\includegraphics[width=1in,height=1.25in,clip,keepaspectratio]{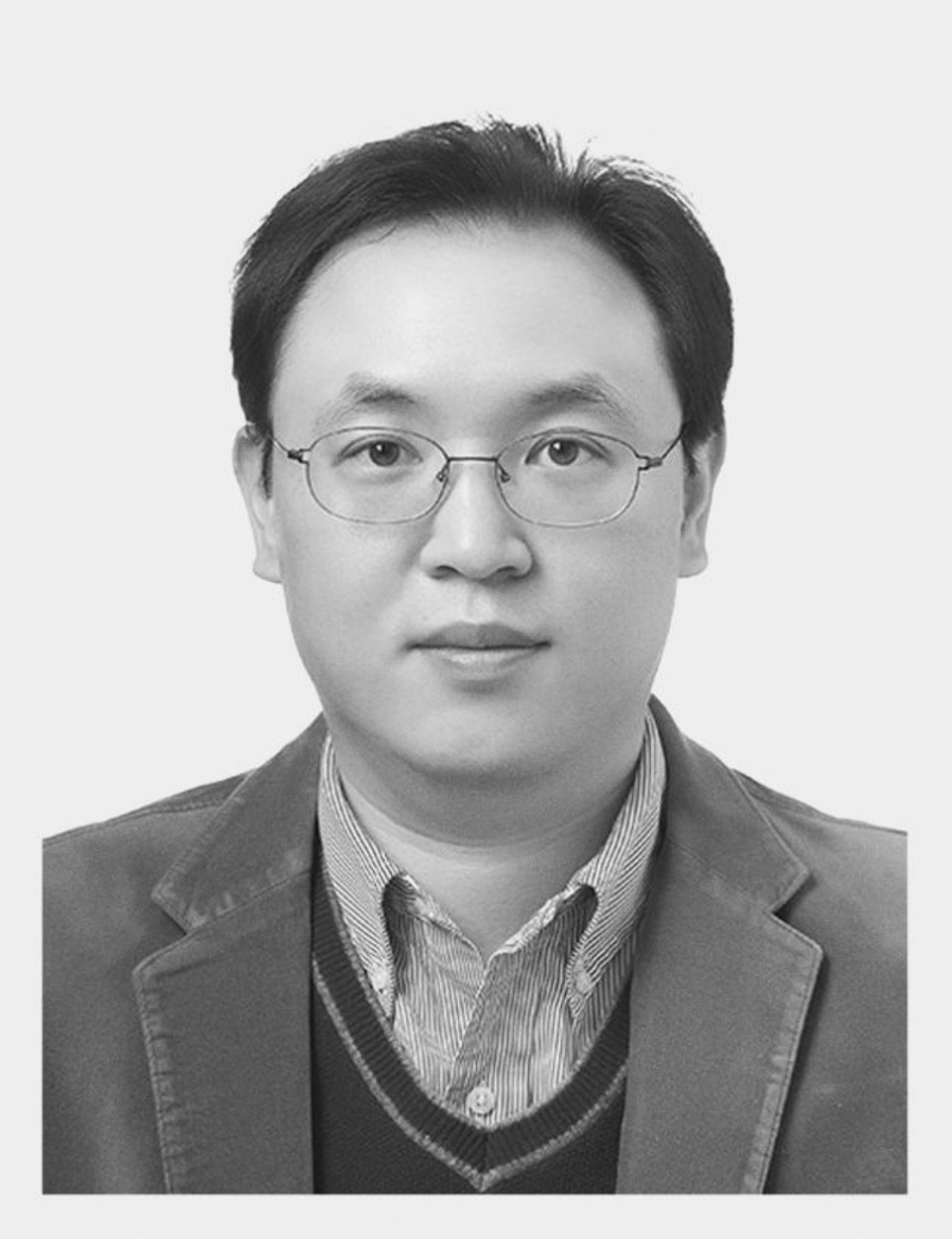}}]{\textbf{Jong Wook Kim}} received his Ph.D. degree from the Computer Science Department, Arizona State University, in 2009. He was a Software Engineer with the Query Optimization Group, Teradata, from 2010 to 2013. He is currently an Assistant Professor of computer science with Sangmyung University. His current research interests include data privacy, distributed databases, and query optimization.
\end{IEEEbiography}

\begin{IEEEbiography}[{\includegraphics[width=1in,height=1.25in,clip,keepaspectratio]{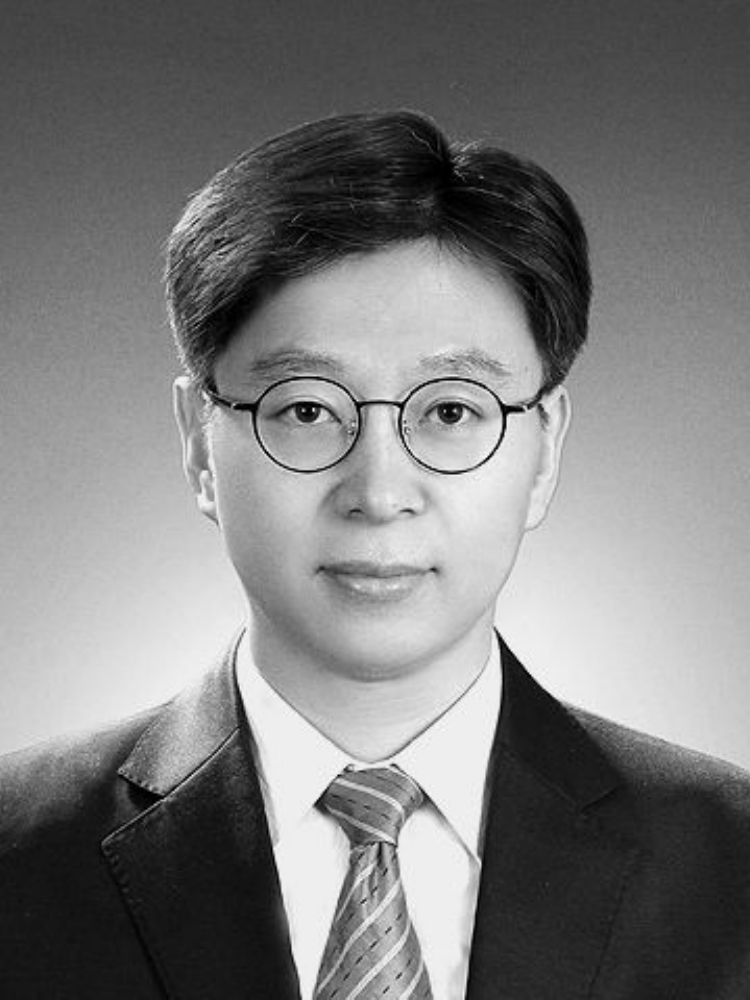}}]{\textbf{Yon Dohn Chung}} received his B.S. degree in computer science from Korea University, and his M.S. and Ph.D degrees in computer science from Korea Advanced Institute of Science and Technology. In 2006, he joined the faculty of the Department of Computer Science and Engineering, Korea University. His research interests include data privacy, spatial databases, and array databases.
\end{IEEEbiography}

\end{document}